\documentclass[twocolumn,prd,nofootinbib,aps,floats,floatfix,amsmath,amssymb,secnumarabic]{revtex4} %
\usepackage[final]{graphicx}
\usepackage{amsmath}
\usepackage{bbm}
\usepackage{amsfonts}
\usepackage{amssymb}
\usepackage{latexsym}
\usepackage{graphicx}
\usepackage[english]{babel}
\usepackage{multirow}
\usepackage{float}
\usepackage{url}
\usepackage{hyperref}
\usepackage{slashed}
\usepackage{xcolor} 

\newcommand{\be}{\begin{equation}}
\newcommand{\ee}{\end{equation}}
\newcommand{\ba}{\begin{array}}
\newcommand{\ea}{\end{array}}
\newcommand{\bea}{\begin{eqnarray}}
\newcommand{\eea}{\end{eqnarray}}

\renewcommand{\d}{\mathrm{d}}

\renewcommand{\d}{\mathrm{d}}

\begin{document}
\preprint{MIT-CTP/4650}
\title{Multimediator models for the galactic center gamma ray excess}
\author{James M. Cline}
\author{Grace Dupuis}
\affiliation{Department of Physics, McGill University,
3600 Rue University, Montr\'eal, Qu\'ebec, Canada H3A 2T8}
\author{Zuowei Liu}
\affiliation{Center for High Energy Physics, Tsinghua University, Beijing, 100084, China}
\author{Wei Xue}
\affiliation{Center for Theoretical Physics, Massachusetts
Institute of Technology, Cambridge, MA 02139, USA}

\begin{abstract} 

Tentative evidence for excess GeV-scale gamma rays from the galactic
center has been corroborated by several groups, including the Fermi
collaboration, on whose data the observation is based.  Dark matter
annihilation into standard model particles has been shown to give a 
good fit to the signal for a variety of final state particles, but
generic models are inconsistent with constraints from direct
detection.  Models where the dark matter annihilates to mediators that
subsequently decay are less constrained.  We perform global fits of
such models to recent data,
allowing branching fractions to all
possible fermionic final states to vary.  The best fit models, including
constraints from the AMS-02 experiment (and also antiproton ratio), require branching primarily to
muons, with a $\sim 10-20\%$ admixture of $b$ quarks, and no other species.  
This suggests models in which there are two scalar mediators that mix
with the Higgs, and have masses consistent with such a decay pattern.
The scalar that decays to $\mu^+\mu^-$ must therefore be lighter than $2m_\tau \cong
3.6$ GeV.  Such a small mass can cause Sommerfeld enhancement, which
is useful to explain why the best-fit annihilation cross section is
larger than the value needed for a thermal relic density.  For light
mediator masses $(0.2-2)$ GeV, it can also
naturally lead to elastic DM self-interactions at the right level for
addressing discrepancies in small structure formation as predicted by
collisionless cold dark matter.

\end{abstract}
\maketitle

\section{Introduction}

Indirect detection may offer the best hope of discovering 
the nature of dark matter beyond its generic gravitational effects.
Currently there is a strong hint from Fermi Large Area Telescope data 
of dark matter annihilation in the
galactic center (GC), giving rise to an excess of gamma rays peaking
at energies of several GeV \cite{Hooper:2010mq,Hooper:2011ti,
Abazajian:2012pn,Zhou:2014lva,Daylan:2014rsa,Calore:2014xka}.
The morphology is consistent with that expected from dark matter
annihilations, and the required cross section is of the right order of
magnitude for yielding the correct thermal relic density.  Millisecond
pulsars are the main conventional astrophysical explanation that has
been proposed \cite{Abazajian:2010zy,Abazajian:2014fta,Yuan:2014rca,
Petrovic:2014xra,Yuan:2014yda} (see however refs.\ 
\cite{Petrovic:2014uda,Carlson:2014cwa} for alternative explanations).
Arguments against the pulsar explanation have been given
in refs.\ \cite{Hooper:2013nhl,Cholis:2014lta,Calore:2014pca}.  While the debate
continues, the dark matter explanation remains a possibility that is
still being widely explored \cite{Anchordoqui:2013pta}-\cite{Bi:2015qva}.

The simplest models have heavy $s$-channel mediators leading to
annihilations $\chi\chi\to f\bar f$ where $f$ represents standard
model particles that lead to gamma rays through their decays (or
possibly inverse Compton scattering).  However even if the mediator
does not couple directly to light quarks, its coupling to any charged
particle induces mixing with the photon at one loop, leading to dark
matter scattering on nucleons.  In many cases the induced coupling
exceeds that allowed by direct detection searches, when the
annihilation cross section for $\chi\chi\to f\bar f$ is large enough
to explain the GC gamma-ray excess.  Cosmic ray antiproton data
also put pressure on direct annihilation to $b\bar b$ 
\cite{Bringmann:2014lpa,Cirelli:2014lwa}
(see however \cite{Hooper:2014ysa}).

To escape direct detection and antiproton constraints, one can consider the
possibility of heavy mediators $\phi$ that are produced on shell
in the annihilations $\chi\chi\to\phi\phi$, and subsequently decay.
Even relatively slow decays due to very small couplings of $\phi$ to
standard model particles can be consistent with the morphology of the 
GC signal, while making the models compatible with direct searches
\cite{Hooper:2012cw}-\cite{Rajaraman:2015xka}.

In the present study we reconsider light mediator models, motivated by
new data sets for the GC excess, one from ref.\ 
\cite{Calore:2014xka} (hereafter CCW) and the other presented by the Fermi
collaboration itself \cite{Murgia}.  These results allow for heavier
dark matter producing gamma rays up to higher energies $\sim 100$
GeV than the earlier
determinations of \cite{Abazajian:2012pn,Daylan:2014rsa} which 
preferred lighter $\sim 30$ GeV dark matter, though the error bars are
large enough for overlap of the allowed parameters.  In addition we 
take into account constraints on annihilations producing electrons,
from the AMS-02 experiment \cite{Aguilar:2014mma,Accardo:2014lma}.  
These turn out to strongly disfavor direct annihilations into
electrons that would produce a bump-like feature in the electron
spectrum.

We find that the CCW and Fermi data show a strong preference for 
annihilation into mediators that decay with branching fractions $\sim
80-90\%$ to muons and  $\sim 10-20\%$ to $b$-quarks.  From a
theoretical model building perspective this at first looks peculiar; 
for a single mediator to have such couplings is not suggested by any
symmetry principle.  On the other hand, singlet scalars that mix with
the Higgs boson will decay preferentially into the heaviest possible standard model
particle, since the couplings are just proportional to those of the 
Higgs.  Therefore a more natural explanation is to have two mediators,
one whose mass is between $2 m_\mu$ and $2m_\tau$, and the other with 
mass between $2 m_b$ and $2 m_t$.  (For a mediator with mass greater
than $2m_h$ one must also consider decays into Higgs bosons.)  
By this combination of fitting to data and theoretical motivations,
we are led to consider models with two scalar mediators with a
hierarchy of masses.  Remarkably, the best fit to the data puts the
mediator masses in the ranges described above, which need not have
been the case.  This provides further experimental motivation for 
building models with several mediators, which at first might have
seemed like a large theoretical leap.

We begin our analysis with an agnostic view concerning theoretical
models,  assuming only a single mediator for simplicity, and allowing
for arbitrary branching ratios into different lepton and quark
flavors, whose spectra are provided by ref.\ \cite{{Cirelli:2010xx}}
(hereafter PPPC).  We describe the construction of the $\chi^2$
functions for fitting to the CCW and Fermi data, respectively,
combined with those of AMS-02.  The preference for final states
consisting of an admixture of muons and $b$ quarks is thereby
established.  We then focus on these final states, adding a separate
mediator coupled to each one, and refit to the data.  In the following
section, a simple model of two scalars mixing with the Higgs is
presented.  We subsequently check whether  the preferred models are
consistent with complementary constraints from direct detection,
thermal relic density, cosmic microwave background, dwarf satellite
observations, the antiproton flux ratio,  and dark matter
self-interactions.  A simple two-mediator model is constructed that
can fit the observations without excessive fine-tuning; Sommerfeld
enhancement of the annihilation in the galaxy plays an important
role.  The model is shown to be consistent with LHC constraints on
extra scalars mixing with the Higgs boson.

\section{Predicted signal}
\label{pred}

The photon flux (GeV/cm$^2$/s/sr) from annihilation of Majorana DM
particles in a given region around the GC can be expressed as
\begin{equation}
E_\gamma^2 {dN^\text{th} \over dE_\gamma} (E_\gamma) =  {1\over
2}\cdot {\bar{J}
\langle \sigma v \rangle\over 4 \pi m_\chi^2}    
\sum_f \text{BR}_{\phi \to f \bar{f}}\, E_\gamma^2 {dN^f \over dE_\gamma}
\label{predflux}
\end{equation}
where $\langle\sigma v\rangle$ is the averaged cross section for DM
annihilation into two mediators, $\text{BR}_{\phi \to f \bar{f}}$ is
the branching ratio for decays of the mediator $\phi$ into $f\bar f$
final states, and $\bar{J}$ is the averaged $J$ factor for the
region of interest (ROI), 
\begin{widetext}
\begin{equation}
\bar{J} = {\int_\text{ROI} d \Omega\, J(l,b) \over \int_\text{ROI} d \Omega }
= {\int_\text{ROI}  \cos(b)db\ d\ell \int_0^\infty dx\, \rho^2
\left(\sqrt{x^2+R_\odot^2-2xR_\odot \cos(\ell)\cos(b)}\right) \over \int_\text{ROI} \cos(b)db\ d\ell }
\label{Jfact}
\end{equation}
\end{widetext}
For Dirac DM, one should replace the prefactor $1/2\to 1/4$ in
eq.\ (\ref{predflux}), resulting
in a cross section that is twice as large as that for Majorana DM,
for a fixed observed flux.
Unless otherwise stated we will assume Majorana DM in the following.
The $J$
factor depends upon the assumed shape of the DM density profile, which is
commonly taken to be of the generalized NFW type,
\begin{equation}
\rho(r) = \rho_s {(r/r_s)^{-\gamma} \over (1+r/r_s)^{3-\gamma}}
\end{equation}
Here we take $r_s=20$ kpc and $\rho_s$ such that 
$\rho_\odot=0.4$ GeV/cm$^3$ in the solar neighborhood, 
with $R_\odot=8.5$ kpc.   Ref.\ \cite{Daylan:2014rsa} finds that the morphology
of the GC excess is best fit by taking $\gamma=1.26$, while ref.\
\cite{Calore:2014xka} adopts $\gamma=1.2$.

${dN^f /dE_\gamma}$ is the photon spectrum from a single DM annihilation into $f\bar f$
final states, and comes from boosting the spectrum due to decay of
the mediators $\phi$ \cite{Agrawal:2014oha}:
\be
	{dN^f \over dE_\gamma} = {2\over (x_+-x_-)}
	\int_{E_\gamma x_-}^{E_\gamma x_+}
	{dE_\gamma'\over E_\gamma'} {dN^f_0 \over dE_\gamma'}
\label{boost}
\ee
where $dN^f_0/dE_\gamma$ is the photon spectrum from $\phi\to f\bar f$ 
in the rest frame of  $\phi$ and $x_\pm = m_\chi/m_\phi\pm
\sqrt{(m_\chi/m_\phi)^2-1}$.  For most final states, a Monte Carlo
generator is needed to predict the distribution of photons from
hadronization and decays.  We take the spectra $dN^f_0/dE$ from
PPPC (ref.\ \cite{Cirelli:2010xx}), which is valid for mediator
masses down to 10 GeV.  For lighter mediators, extrapolation of 
the PPPC results would be required, introducing inaccuracies into
the predictions.  However our subsequent fits will dictate the need
only for lighter mediators decaying into muons, for which we use an
analytic expression for $dN^\mu_0/dE_\gamma$ \cite{Mardon:2009rc}, given in
appendix \ref{appA}.  The factor of 2 in (\ref{boost}) accounts for
the two mediators produced in the annihilation.

\begin{figure}[t]
\includegraphics[width=\columnwidth]{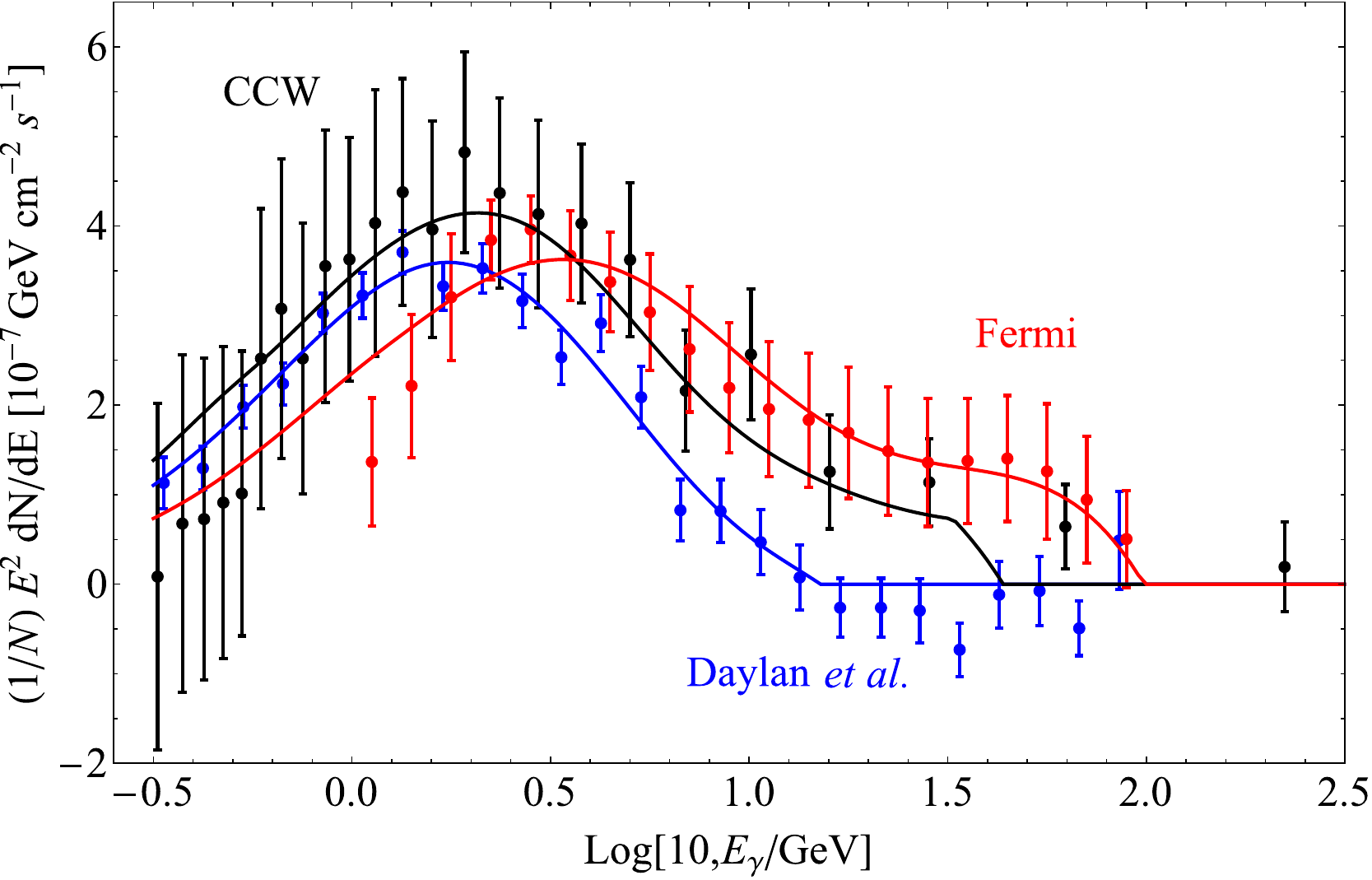}
\caption{The three data sets for the GC gamma ray excess.
The Fermi flux presented is the total flux from the
$15^\circ\times 15^\circ$ square
around the GC; the other two fluxes are normalised accordingly with same DM profile.
Error bars for CCW are taken from the diagonal components
of their covariance matrix.  Solid curves are the predictions
of the best-fit models described in section \ref{1medfits}.
}
\label{fig:data}
\end{figure}

\section{Data sets}

We consider three data sets for the GC excess: those of CCW \cite{Calore:2014xka},
Fermi \cite{Murgia}, and Daylan {\it et al.} \cite{Daylan:2014rsa}.  Detailed
descriptions follow; the data are summarized in fig.\ \ref{fig:data}.
We further include the AMS-02 measurement of the 
spectrum of cosmic ray electrons/positrons, and antiproton ratio
data from the BESS, CAPRICE and PAMELA experiments.  The antiproton
data is not included in our overall $\chi^2$ function; rather we will
verify after fitting the data that our models are compatible with
the antiproton constraints.

\subsection{CCW spectrum}

The CCW spectrum (ref.\
\cite{Calore:2014xka}) is downloadable from \cite{CCW}, where a
covariance matrix $\Sigma$ for computing the $\chi^2$ for fits to the data is
also provided.  Then
\bea
\chi^2 &=& \sum_{ij}^{24} \left( E_i^2 {dN^\text{th}\over dE}(E_i) - E_i^2 {dN^\text{exp}\over dE}(E_i) \right)
(\Sigma^{-1})_{ij}\cdot\nonumber\\
&&\left( E_j^2 {dN^\text{th}\over dE}(E_j) - E_j^2 {dN^\text{exp}\over dE}(E_j) \right)
\label{CCWchi2}
\eea
where the sum is over 24 energy bins, and $(\Sigma^{-1})_{ij}$ are the
matrix elements of the inverse of the covariance matrix,
obtained from 
columns 29-52 of the data file from \cite{CCW}).  This accounts for
the correlations between the different energy bins.

The ROI for CCW is a $\pm 20^\circ$ square around the GC;
 in galactic coordinates, where $\ell$ is the longitude and $b$ is the latitude,
the region is  
\begin{equation}
|\ell|<20^\circ ~\text{and}~ 2^\circ<|b|<20^\circ
\end{equation}
where the $\pm 2^\circ$ region in latitude is masked out to remove the
galactic disk.

With these inputs and $\gamma=1.2$ for the generalized NFW
distribution, the $J$ factor
in (\ref{predflux})  is $\bar{J}=2.062 \times 10^{23}\ \text{GeV}^2\ \text{cm}^{-5}$ 
for the CCW data. 
(This number agrees with \cite{Agrawal:2014oha}.)  

\begin{figure}[t]
\includegraphics[width=1\columnwidth]{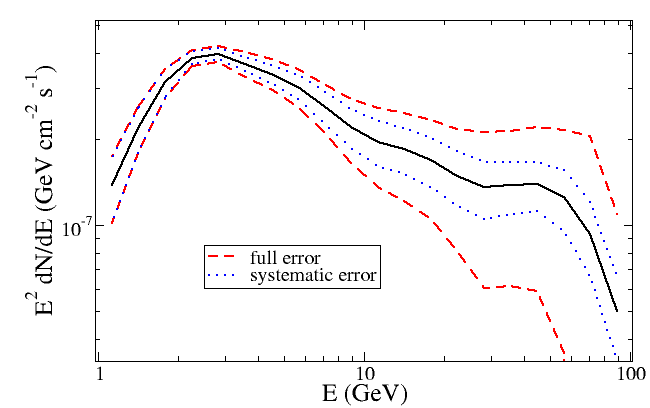}
\caption{Spectrum for GeV excess from Fermi data, extracted from
ref.\ \cite{Murgia}}
\label{fig:fermi-flux}
\end{figure}

\subsection{Fermi spectrum}
\label{fermi-spect}

The Fermi collaboration has not officially released its data, but we
have digitized it from the presentation in ref.\ \cite{Murgia},
and list the results in table \ref{table:fermi}, appendix \ref{App0}.
Fluxes are given in 20 energy bins, equally spaced in $\log_{10}(E)$
between 1 MeV and 89 GeV.   We estimate the statistical errors from
taking $\sqrt{N}$ for the number of total events in each bin and
applying this to the part of the signal interpreted as the excess.
Ref.\ \cite{Murgia} gives two characterizations of the spectrum of
excess events, one which is presumed to be a power law in energy with
an exponential cutoff, and the other being a separate fit to the
excess in each bin.  We adopt the latter for our analysis.  

In
addition to the statistical errors, there is systematic uncertainty 
associated with assumptions about the templates for background photons
from pulsars and OB stars.  We define the signal as
the median between the upper and lower envelopes found from varying
these templates, and the systematic error as the difference.  This
is added in quadrature with the statistical error to estimate the
total uncertainty.  The result is plotted in fig.\
\ref{fig:fermi-flux}, showing that errors are systematics-dominated at
low energy, but mostly statistical at high energy.
The $\chi^2$ function is then defined in the usual
way.

The ROI for the Fermi data is a $15^\circ\times 15^\circ$ square
around the GC.  Numerically integrating (\ref{Jfact}) we obtain
$J = 1.07\times 10^{23}$ GeV$^2$/cm$^5$, again assuming 
generalized NFW parameter $\gamma=1.2$ and $\rho_\odot =
0.4\,$GeV/cm$^3$.  Here we do not average over the solid angle
(giving $J$ instead of $\bar J$) because the Fermi data are reported
as a total flux rather than an intensity flux.

\begin{figure}[t]
\includegraphics[width=\columnwidth]{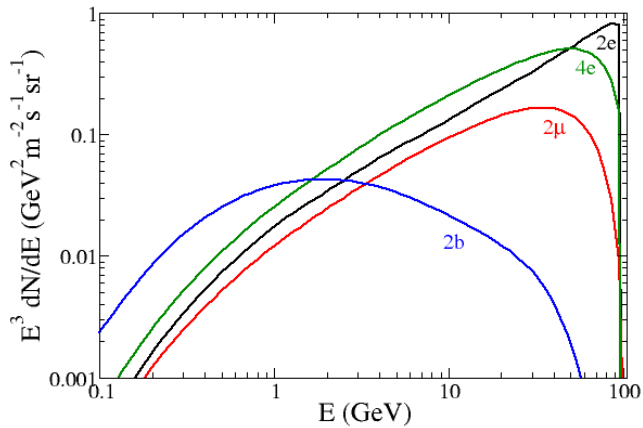}
\caption{Electron spectra from dark matter annihilation into
$2e$, $4e$, $2\mu$ and $2b$, showing why annihilations directly into 
electrons are most strongly constrained by AMS-02 data.}
\label{fig:ams-spect}
\end{figure}

\begin{figure*}[t]
\includegraphics[width=2\columnwidth]{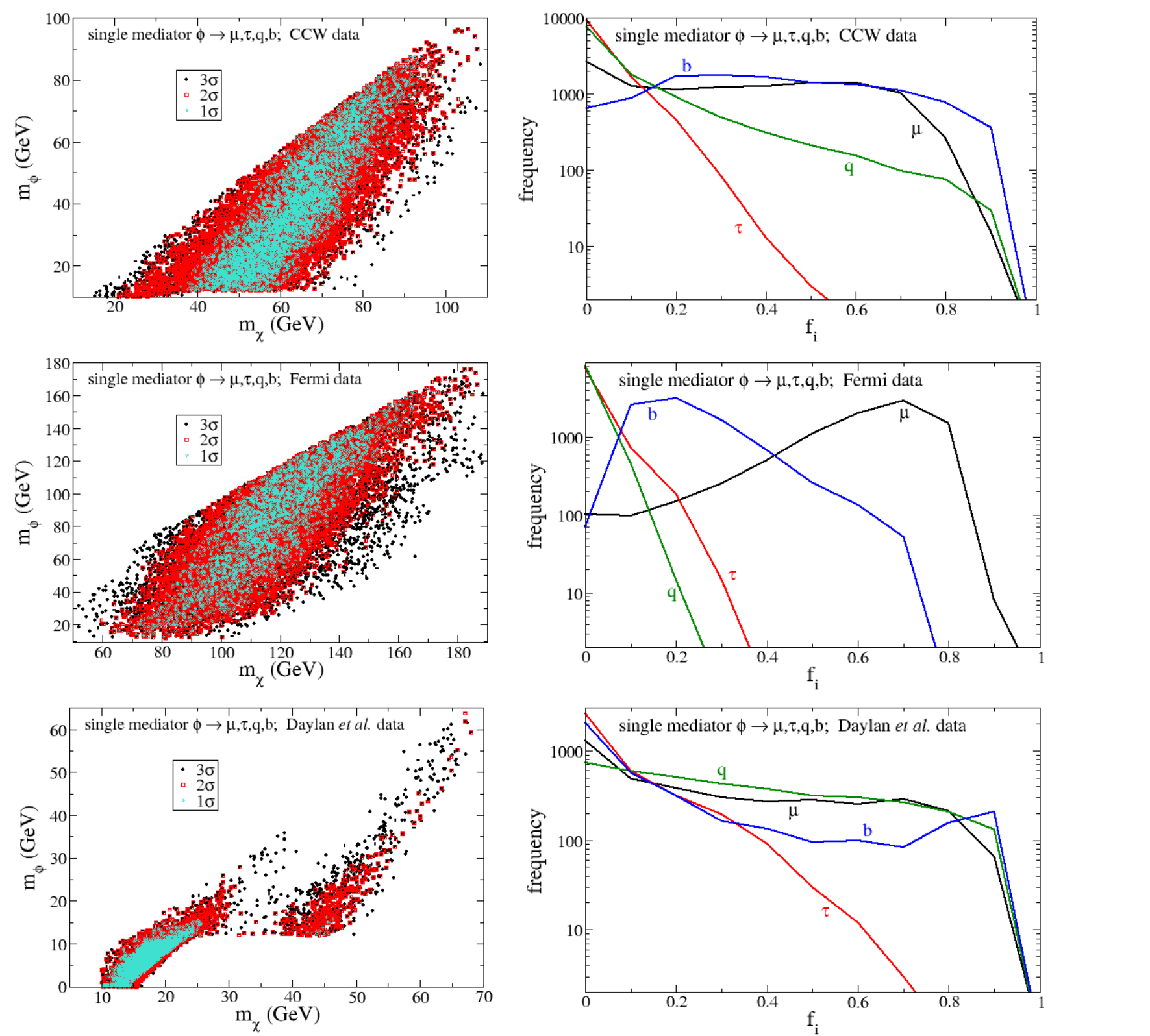}
\caption{Left column: distribution of $m_\phi$ versus $m_\chi$ for single-mediator
models from fitting to CCW, Fermi and Daylan {\it et al.} data (from
top to bottom), with $f_e=0$ and floating branching
fractions to $\mu,\tau,q,b$.  Right column: corresponding
distributions of $f_\mu,f_\tau,f_q,f_b$.}
\label{fig:1med-panel}
\end{figure*}

\subsection{Daylan {\it et al.} spectrum}

Although our primary purpose is to explore the implications of the 
first two data sets which are more recent, for completeness we also
apply our methodology to the GC excess spectrum determined
in ref.\ \cite{Daylan:2014rsa}.  It can be read from fig.\ 5 of that
paper, from which it is straightforward to define $\chi^2$ for a given
predicted flux.  The spectrum shown there has been normalized to
correspond to a $J$ factor that is not averaged over any solid angle,
but instead is defined along a line of sight $5^\circ$ away from the
GC.  For $\gamma = 1.2$, we thus find $J = 9.09\times 10^{23}$
GeV$^2$/cm$^5$ (again using  $\rho_\odot = 0.4\,$GeV/cm$^3$).

\subsection{AMS positron/electron spectrum}

AMS-02 published their recent results on the measurement of  the
$e^+/e^-$ ratio and the separate electron and positron fluxes in
\cite{Aguilar:2014mma,Accardo:2014lma}, confirming a positron excess
above  the expectations from standard cosmic ray propagation scenarios
that was previously seen by PAMELA and Fermi. The more precise
measurements of AMS-02 do not tell us whether the excess of the
positron flux originates from dark matter or pulsars, but the smooth
spectrum up to $\mathcal{O} (100)\,$GeV already puts strong
constraints on dark matter models
\cite{Bergstrom:2013jra,Hooper:2012gq,Ibarra:2013zia,Liu:2014cma}.  If
light dark matter annihilates directly into $e^+ e^-$, even though the
spectrum is initially a delta function of energy,  after propagation
it leads to  bump-like feature in the observed spectra.   The feature
is more spread out in models with decays into mediators,
 $\chi\chi\to\phi\phi\to 4e$, where
the initial spectrum is box-like; nevertheless the final shape is
still localized in energy and can be strongly constrained.
The differences in shape of the electron spectra for relevant final
states are illustrated in fig.\ \ref{fig:ams-spect}.

To simplify our analysis and to make the results more
model-independent,  we assume that only dark matter annihilating to
$e^+ e^-$ or $2 e^+ 2 e^-$ can be constrained by current AMS-02 data,
since these have harder and more localized spectra 
compared to other channels.  The latter 
also generate positrons and electrons, but they come from 
three- or four-body final states, which make the
spectrum much softer and more broad.  It would be easy to absorb 
such nondistinctive contributions into the smooth 
background.\footnote{The background may include contributions from 
primary and secondary electrons, secondary positrons, pulsars, or 
heavy dark matter annihilation/decay. All of them could contribute to the 
smooth spectra observed by AMS-02.}  Following
\cite{Liu:2014cma}, we use polynomial functions to fit the  logarithm
of AMS spectra as the background.  To obtain the electron/positron
flux from light dark matter, a cosmic ray  propagation model is 
chosen
with a relatively large magnetic field,  in order to achieve 
conservative bounds,  since the energy loss is essentially due to the
magnetic field and will soften the spectra.

\begin{figure}[t]
\includegraphics[width=\columnwidth]{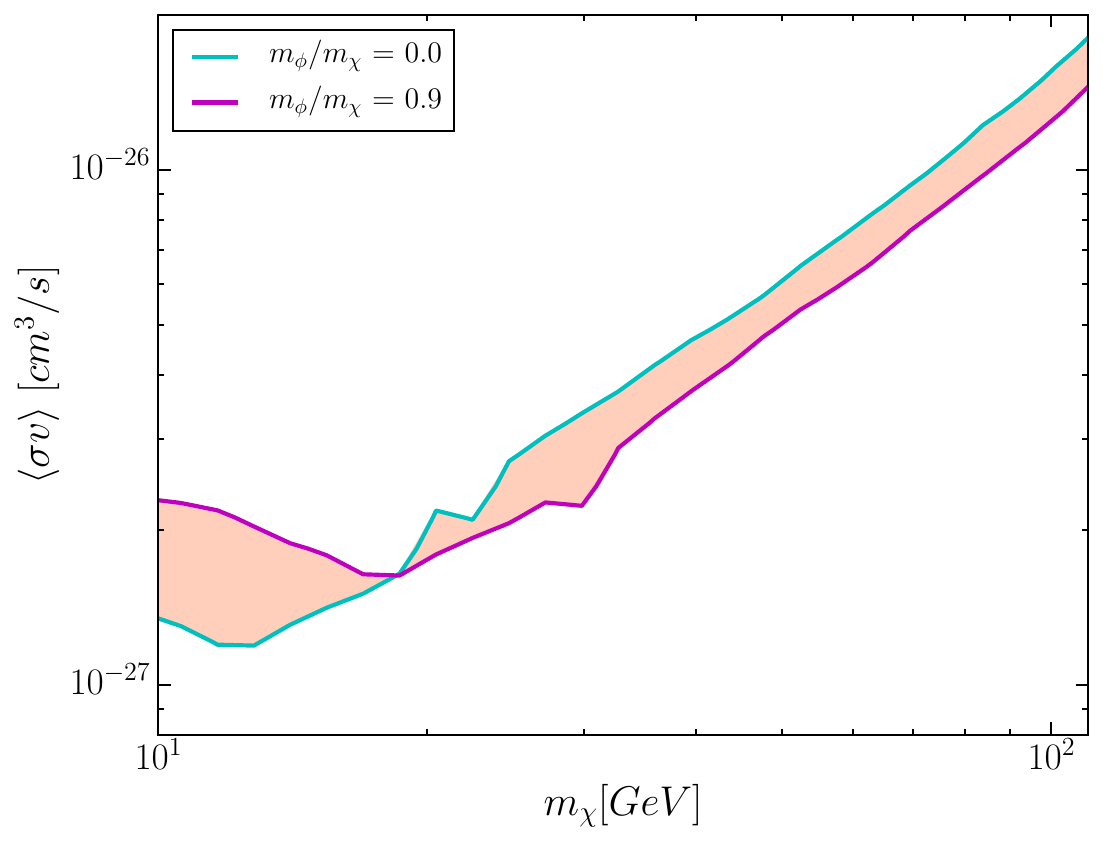}
\caption{AMS exclusion curves for 
$\chi + \chi \rightarrow 2 \phi \rightarrow 4 e$.
The exclusion limit for two values of $m_\phi / m_\chi = 0$ and 0.9
are shown here, to illustrate the (relatively weak) dependence on 
the mediator mass.
}
\label{fig:AMSexclusion}
\end{figure}

Before including the light dark matter contribution, the absolute
$\chi^2 $ of the background to fit against the AMS positron ratio data
is $\chi^2 = 35.06$. This value does  not depend on assumptions about
the propagation model or solar modulations since we use fitting
functions, rather than considering specific propagation models. 
However for the dark matter signal, the propagation and solar
modulation uncertainties have to be considered. We marginalize over
the range of the effective potential of solar
modulation, $[0, 1\, \mathrm{GV}]$ and a relatively
large magnetic field is chosen to propagate positrons and electrons
from dark matter annihilation.\footnote{The total magnetic field is composed of
a regular one and a turbulent one. We normalize the total value at the Sun to $B_\odot = 
15 \mathrm{\mu G}$.} The 3$\sigma$ exclusion curves for
dark matter annihilating to $2e^+ 2e^-$ are illustrated in fig.\
\ref{fig:AMSexclusion} for two different mass ratios  $m_\phi/m_\chi$. The
strong constraints from AMS lead us to models with negligible direct
annihilations into electrons.

\begin{figure}[t]
\includegraphics[width=1\columnwidth]{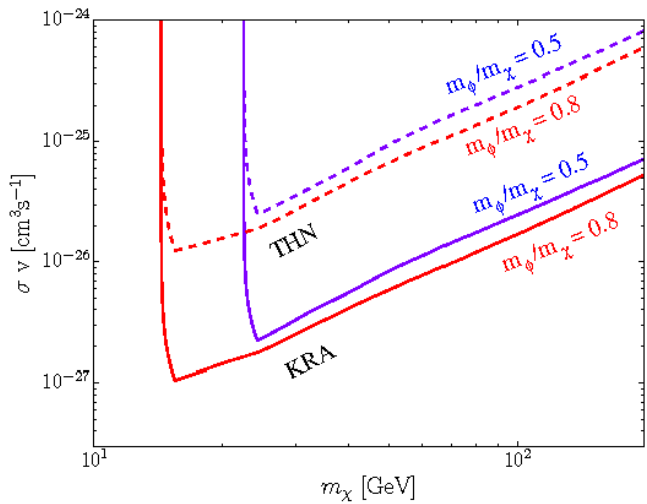}
\caption{
Antiproton exclusion curves for $\chi+ \chi \rightarrow 2 \phi 
\rightarrow b b \bar{b} \bar{b}$. The $3\sigma$ exclusion limit for two propagation 
models, KRA and THN, and for two values of $m_\phi / m_\chi = 0.5$ and
$m_\phi/m_\chi = 0.8$ are shown here respectively.
}
\label{fig:AntiProton}
\end{figure}

\begin{figure*}[t]
\centerline{\includegraphics[width=1.2\columnwidth]{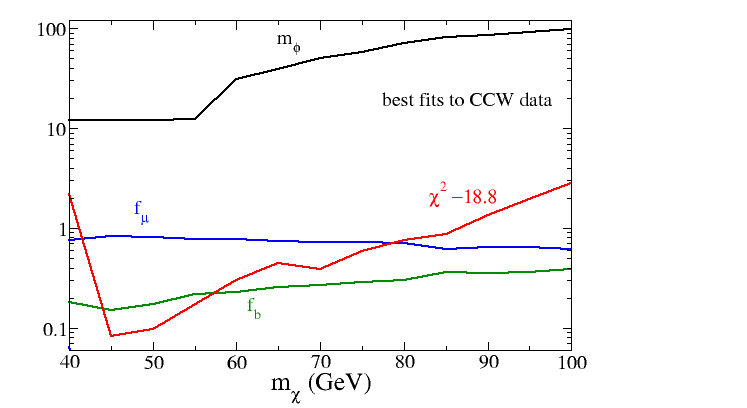}
$\!\!\!\!\!\!\!\!\!\!\!\!\!\!\!\!\!\!\!\!\!\!\!\!\!\!\!\!\!\!\!\!\!$\includegraphics[width=1.2\columnwidth]{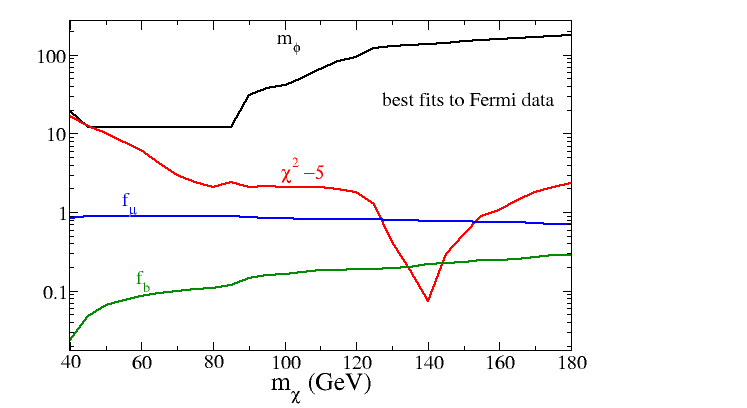}}
\centerline{$\!\!\!\!\!\!\!\!\!\!\!\!\!\!\!$\includegraphics[width=1.2\columnwidth]{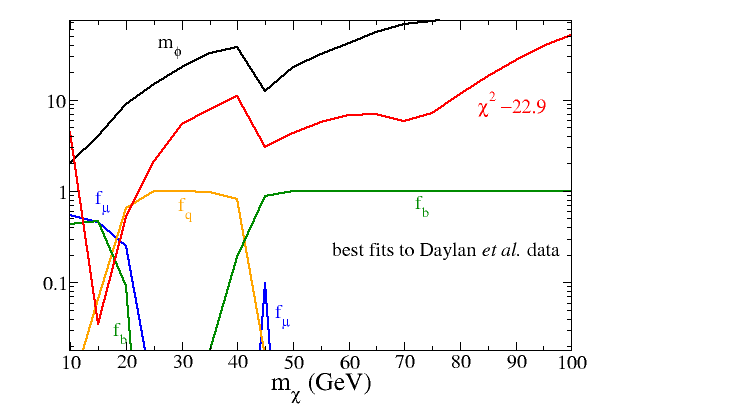}
$\!\!\!\!\!\!\!\!\!\!\!\!\!\!\!\!\!\!\!\!\!\!\!\!\!\!\!\!\!\!\!\!\!$
$\!\!\!\!\!\!\!\!\!\!\!\!\!\!\!\!$
\includegraphics[width=1.14\columnwidth]{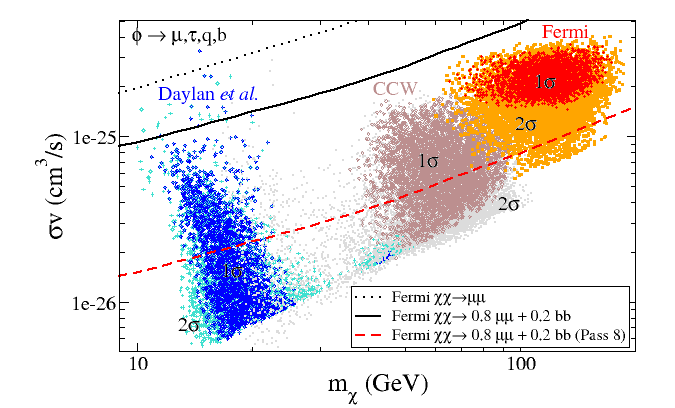}}
\caption{Top row and bottom left: best-fit values of $m_\phi$ (in GeV), $f_\mu$ and $f_b$ as a function of 
$m_\chi$ for data from CCW (top left), Fermi (top right) and Daylan {\it et
al.} (bottom left).  Bottom right: distributions of  cross sections 
versus $m_\chi$ (1-2$\sigma$ allowed regions) for
the three data sets, compared to estimated Fermi limits from $\chi\chi$
annihilation to $\mu\mu+bb$ in dwarf satellites (see text). }
\label{fig:degen}
\end{figure*}

\subsection{Antiproton ratio}

The current measurements of the cosmic ray antiproton flux agree well
with the expected astrophysical backgrounds; therefore, they are able
to significantly constrain the cross section  for dark matter
annihilation to hadronic final states, modulo the uncertainties from
cosmic ray propagation. Following \cite{Urbano:2014hda}, we use
antiproton ratio data from the
BESS~\cite{Orito:1999re,Asaoka:2001fv},  CAPRICE~\cite{Boezio:2001ac}
and PAMELA~\cite{Adriani:2012paa} experiments. In order to minimize the
systematic uncertainties from different experiments, the antiproton
ratio data are employed here, rather than the antiproton flux data. 
Parameters governing cosmic ray propagation are constrained by
fitting to the observed ratio of boron to carbon, but this still
leaves some freedom to vary parameters, as well as the solar
modulation;
this leads to a range of 
estimates for the astrophysical background of proton and antiproton
fluxes, that are consistent with the observed fluxes from the above
data sets.  
 Once
the propagation model, the solar modulation and the astrophysical 
background are fixed, we can perform a $\chi^2$ fit of the $\bar{p}/
p$ data by adding the contribution from the dark matter annihilation.

Two benchmark  propagation models are used here to cover the
systematic uncertainties of  the cosmic ray propagation: KRA and THN,
in the notation of ref.\ \cite{Evoli:2011id},  where the main
difference  of the two models is the halo height, with $z_t = 4\,
\mathrm{kpc}$ for  KRA, and $z_t = 0.5\, \mathrm{kpc}$ for THN. The
other parameters are  the slope of the diffusion coefficient  $\delta
= 0.5$, nuclei spectral index $\gamma = 2.35$, the normalization  of
the diffusion coefficient $D_0 = 2.68 \times 10^{28}\, \mathrm{cm^2
s^{-1}}$, and the solar modulation potential $\Phi = 0.95\,
\mathrm{GV}$ for the KRA model.  For THN these take the same values
except for   $D_0 = 0.32 \times 10^{28}\, \mathrm{cm^2 s^{-1}}$. 
Neither of these models consider convection.  Further details can be
found in  \cite{Urbano:2014hda}. 

The resulting antiproton constraints on DM annihilating to  $bb
\bar{b}\bar{b}$ with two  different ratios of mediator mass to DM mass are
shown in fig.~\ref{fig:AntiProton}. The dependence on propagation
model is the largest uncertainty, with THN (dashed lines)  giving a
factor of 10 weaker constraints than KRA (solid lines), while the
dependence on the mediator mass is relatively weak (red versus blue
curves).  We adopt the more conservative bounds from the THN model
in our analysis.  It will turn out that in our preferred models
with a subdominant branching fraction to $b$ quarks, the antiproton
limit is not significantly constraining (see fig.\ \ref{fig:two-med2} below).

Better measurements of heavy nuclei abundances from AMS-02, in
particular the boron/carbon ratio, may improve our understanding of
cosmic ray propagation models, hopefully leading to a decrease in  the
systematic uncertainty from the propagation models.  Moreover a much
more precise measurement of antiproton data itself is
expected to improve this limit by a factor of two or three.
Considering both effects, we anticipate that coming AMS-02 data may
shed further light on which DM models can consistently explain the GC
excess.

\section{Fits with a single mediator}
\label{1medfits}

Our initial motivation was to reexamine the viability of  kinetically
mixed $Z'$ models for the GC excess  in light of the new data sets
from CCW and Fermi.  Although a good fit to the GC excess can be
obtained, these models necessarily have a significant branching
fraction into electrons.  We found that the combined fit to the GC
excess and AMS electron data was quite poor.

This suggests doing a model-blind search for different combinations
of final state fermions that could give a good fit to all data.
We performed this in the 8-parameter space of models characterized by the
DM and mediator masses $m_\chi$, $m_\phi$, and branching fractions
$f_i$ for annihilation into final states $i = e,\mu,\tau,q,c,b$,
where $q$ denotes light quarks, whose spectra are all provided by
PPPC \cite{Cirelli:2010xx}.  These fractions are subject to the
constraint $\sum_i f_i =1$, where we ignore invisible channels since
this degree of freedom would be degenerate with the 
overall cross section $\langle\sigma v\rangle$, which makes the 8th
parameter.

\begin{table}[t]
\begin{tabular}{|c||c|c|c|c|c|c|c|c|c|c|}
 \hline
   Data set &   $m_\chi$ &   $m_\phi$ &  $f_\mu$ &  $f_\tau$ & $f_q$ &
 $f_b$ &   $\langle\sigma v\rangle$ &  $\chi^2_{\rm min}$ &  DOF  \\
\hline
 CCW  & 46 & 12.3  & 0.82 & 0 & 0.02 & 0.16 & $1.1$ & 18.8
& 	24     \\
 \hline
Fermi & 130& 114.5 & 0.80 & 0 & 0 & 0.20 & $2.8$ & 6.4
& 20 \\
\hline
Ref.\ \cite{Daylan:2014rsa} &14.6  & 4.0 & 0.49 & 0.01 & 0.06 & 0.44 &
$0.7$ & 22.9 & 25\\
 \hline
\end{tabular}
\caption{Best fit parameters for single-mediator model fits to the
three data sets.  Masses are in GeV units, $\langle\sigma v\rangle$ in
units of $10^{-25}\,$cm$^3$/s.  ``DOF'' is the number of data points in each set.}
\label{tab:best-fit}
\end{table}

\begin{table}[t]
\begin{tabular}{|c|c|c|c||c|c|c|c|c|}
 \hline
   Data set &   $m_\chi$ &   $m_\phi$ &  $\chi^2_{\rm min}$ &  
   Data set  & $m_\chi$  &  $m_\phi$  
&$\langle\sigma v\rangle$&  $\chi^2_{\rm min}$  \\
\hline
CCW ($\mu$) &  8.5 & 1 & 37 & CCW ($b$) & 75 & 37 & 0.6 & 23.6 \\ 
 \hline
Fermi ($\mu$) & 14 & 1 & 42 & Fermi ($b$) & 153 & 102 & 1.5 & 20.5 \\
 \hline
Ref.\ \cite{Daylan:2014rsa} ($\mu$) & 7.5 & 1 & 51 & Ref.\
\cite{Daylan:2014rsa} ($b$)
 & 51 & 25 & 0.4 & 27.5 \\
\hline
\end{tabular}
\caption{Best fit parameters for single-mediator model fits to the
three data sets, allowing for 100\% branching to $\mu$ final states
(left) or $b$ (right).
  Masses are in GeV units, $\langle\sigma v\rangle$ in
units of $10^{-25}\,$cm$^3$/s.}
\label{tab:best-fit2}
\end{table}

The result of this search is that $f_e$ must be negligible as a result
of the AMS constraint.  Therefore for more refined searches we set
$f_e=0$ and remove the AMS contribution from the total $\chi^2$.
We find that the CCW and Fermi GC
spectra further disfavor any significant contribution from $\tau$,
$q$ or $c$, preferring an admixture of muons and 
$b$ quarks.  We sample models with $\chi^2$ near the minimum value
using a Markov Chain Monte Carlo (MCMC) search as well as Multinest
\cite{2013arXiv1306.2144F}; 
this prevents getting
stuck in spurious local minima of the $\chi^2$ function.  In this
way we find a large spread in allowed masses 
$m_\chi$ and $m_\phi$ as shown in fig.\ \ref{fig:degen}.
The best fit values of parameters for the three data sets are listed
in table \ref{tab:best-fit}.

However the nominal best-fit values should not be given too much
importance, because there are quasi-degeneracies in the $\chi^2$ functions that
allow for good fits to the data over a range of parameters for the
Fermi and CCW data sets.  These are illustrated in fig.\
\ref{fig:degen}.  We see that $m_\chi$ can vary in the range
40-100 GeV keeping $\Delta\chi^2 \lesssim 2$ for the CCW fits.
Similarly the Fermi data are compatible with $m_\chi$ in the range
80-180 GeV.  Both data sets are compatible with annihilations into
$\mu\bar\mu$ and $b\bar b$, with the former being dominant.
The corresponding results for the Daylan {\it et al.} spectrum are
qualitatively different, both for the lower range of 
$m_\chi$ and the final states, which prefer $\mu$, $q$ or $b$
depending upon $m_\chi$.  A range of larger masses $m_\chi \sim 55-70$
GeV is however still reasonable with $\Delta\chi^2 \sim 6$.  This
provides overlap with the preferred regions of the other data sets.

The required values of the annihilation cross section $\langle\sigma
v\rangle$ needed to fit the magnitude of the excess are illustrated in
fig.\ \ref{fig:degen} (bottom right).  The 1-2$\sigma$  favored 
regions in the $m_\chi$-$\langle\sigma v\rangle$ plane are somewhat
disjoint between the three data sets.  We show for comparison
estimated limits on annihilation to $80\%\,\mu\mu+20\%\,bb$ from Fermi
observations of satellite galaxies of the Milky Way
\cite{Ackermann:2013yva}, with an interpolation to the case of
interest for us of $f_\mu\sim 0.8$, $f_b \sim 0.2$ using
$\langle\sigma v\rangle^{-1}  \cong f_\mu\langle\sigma_\mu
v\rangle^{-1} + f_b\langle\sigma_b v\rangle^{-1}$.  Here the
$\sigma$'s refer to the limiting values.  This is a rough estimate of
the expected limit on our best-fit models to the CCW and Fermi GC
excess data, which is seen to be still consistent with most of the
perferred parameter space.   Also plotted is our estimate of the
limit on $\langle\sigma_{b}v\rangle$ from Fermi's Pass 8
data \cite{Ackermann:2015zua}, which is lower than that of ref.\
\cite{Ackermann:2013yva} by a factor of 5.  These results
indicate some tension between the dwarf limits and the
GC excess signal.  This can be relieved somewhat, still consistently
with our fits, by reducing the fraction $f_b$ so that the signal
is more $\mu$-like.  Fig.\ \ref{fig:degen} (bottom right) shows that
the constraints into pure $\mu$ final states are significantly weaker.

Since the allowed ranges for $m_\chi$ include values greater than
$m_W$ and $m_Z$, we have also performed Monte Carlo searches of the
enlarged parameter space allowing for branching into $W$ and $Z$.
However we find that the best fit models have negligible branching
into these states, and so we henceforth disregard them.

One may wonder whether the preference for an admixture of $\mu$ and
$b$ final states is statistically significant compared to  100\%
branching to either $\mu$ or $b$.  We have scanned over  $m_\chi$ and
$m_\phi$ for these cases to compare with the previous results for
admixtures of final states.  The best-fit parameters and minimum
values of $\chi^2$ are given in table \ref{tab:best-fit2}, to be
compared to the results of table \ref{tab:best-fit}.  Annihilations to
$\mu$ only give poor fits to all data sets, while those to $b$ do
better.  Nevertheless, the improvement from allowing an admixture of
$\mu$ and $b$ is statistically significant.   The $q$ and $\tau$
channels play a negligible role in the fits to multiple final states,
so these can be  considered as approximately the same as having only
mixed $\mu$ and $b$ states. The improvements in $\chi^2$ with two
channels versus $b$ only is $4-5$ for the CCW and ref.\
\cite{Daylan:2014rsa} data sets,  and 14 for the Fermi data.  In all
cases $\delta\chi^2\gg 1$ for the inclusion of one extra parameter.

Physically, the preference for two final states arises because the
spectrum from any one of them falls with energy as a power law times
an exponential cutoff.  Especially for the Fermi data (see fig.\ 
\ref{fig:data}), the observed spectrum does not have this form  but
instead remains high to large energies, just before abruptly cutting
off.  The superposition of final state spectra from $\mu$ and $b$
is able to reproduce this shape much better than that from $b$s alone.

\begin{figure*}[t]
\centerline{
\includegraphics[width=1.2\columnwidth]{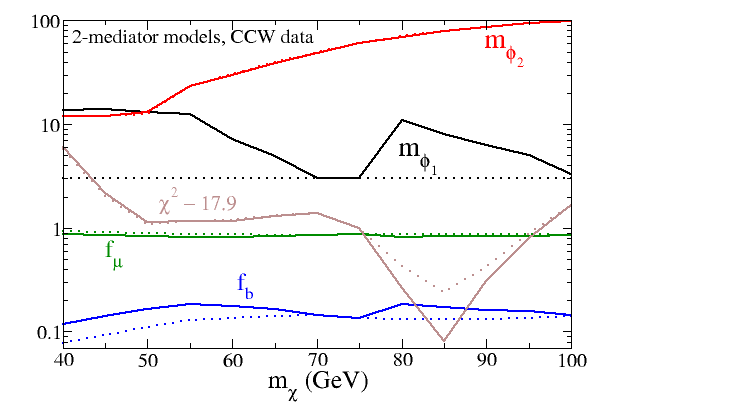}
$\!\!\!\!\!\!\!\!\!\!\!\!\!\!\!\!\!\!\!\!\!\!\!\!\!\!\!\!\!\!\!\!\!$
\includegraphics[width=1.2\columnwidth]{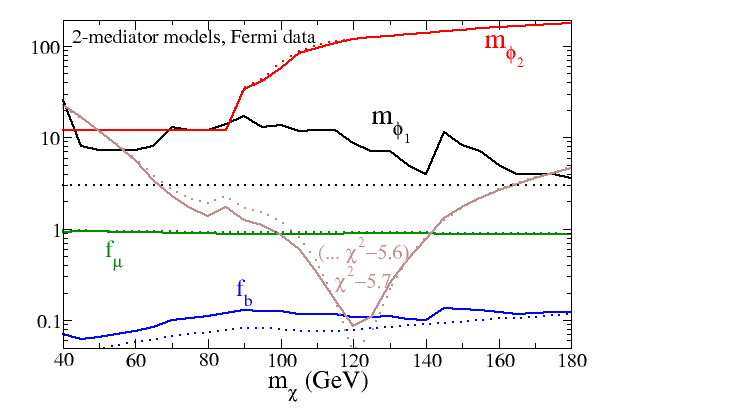}
}
\caption{Left: best fits as a function of $m_\chi$ to the CCW
data, as in fig.\ \ref{fig:degen}.  Solid lines have $m_{\phi_1}$
freely varying, while dotted ones restrict $m_{\phi_1}=3$ GeV.
Right: same for Fermi data.}
\label{fig:two-med}
\end{figure*}

\begin{figure*}[t]
\centerline{
\includegraphics[width=1.1\columnwidth]{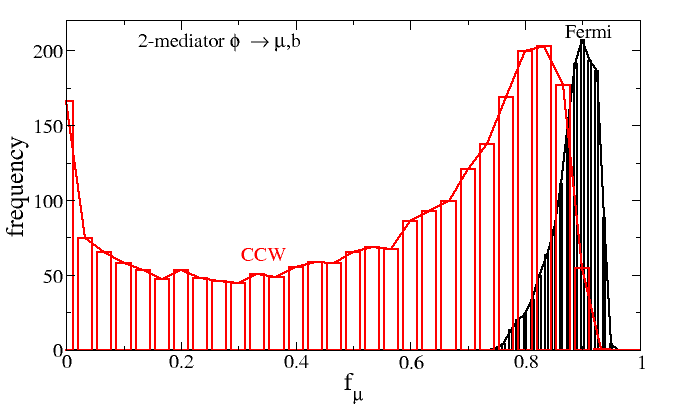}
$\!\!\!\!\!\!\!\!\!\!\!\!\!\!\!\!\!$
\includegraphics[width=1.1\columnwidth]{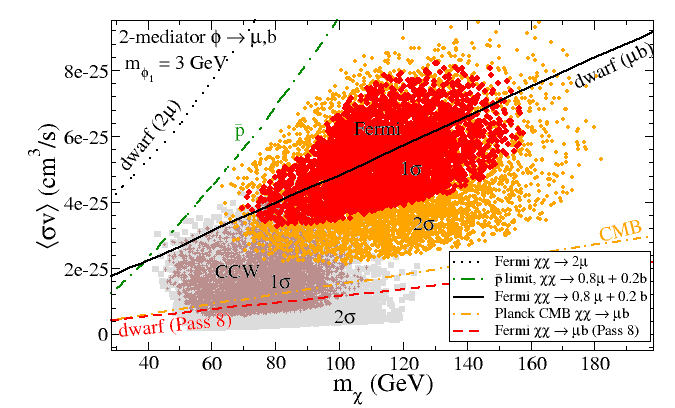}
}
\caption{Left: distribution of branching fraction to muons for
two-mediator model, with $m_{\phi_1}$ fixed at 3 GeV.  Right:
corresponding distributions of $\langle\sigma v\rangle$ versus
$m_\chi$, with Fermi dwarf galaxy constraints overlaid, as well
as our constraint from antiprotons and that from the CMB. }
\label{fig:two-med2}
\end{figure*}

\section{Fits with two mediators; Sommerfeld enhancement}

The previous analysis shows that annihilation to a mixture of
$b$ and $\mu$ final states gives the best fits to Fermi and CCW data.
From a model-building point of view, it may seem ad-hoc to couple the
mediator to the standard model fermions in this way.  If
the mediator is a singlet that mixes with the Higgs, the couplings
are proportional to the masses of the fermions.  With two such 
mediators however, the desired mixture of final states could be
a natural consequence of the masses $m_{\phi_1}$ and $m_{\phi_2}$, only 
requiring that $m_{\phi_1} < 2 m_\tau$ and $2m_b < m_{\phi_2} <
m_\chi$, so that decays into disfavored channels are kinematically
forbidden, or suppressed by small Yukawa couplings.
Motivated by this theoretical consideration and the previous
results, we thus
reconsider the data in the model with five parameters 
\be
\{m_\chi,\, m_{\phi_1},\, m_{\phi_2},\, f_b/f_\mu,\, \langle\sigma v\rangle\}
\nonumber
\ee
This is one fewer parameter than in the single-mediator models, where
we had two additional final states.  

Interestingly, the fits to Fermi and CCW data, in which $m_{\phi_1}$ is
free to vary, are consistent with values that are not far from the theoretical
threshold $2m_\tau$.  In fig.\  \ref{fig:two-med} we plot best-fit
parameters that are at least local minima of $\chi^2$, demonstrating
this assertion.  $\chi^2$ is sufficiently flat as a function of
$m_{\phi_1}$ that the models remain good fits even when
$m_{\phi_1}$ is restricted to stay below $2m_\tau$, as is also shown
in the figure.  Monte Carlo searches of the parameter space reveal
slightly better fits ($\delta\chi^2\sim 0.1-0.2$) 
with higher $m_{\phi_1}$, but these do not contradict the goodness of
fit of the models with $m_{\phi_1}< 2m_\tau$. 

Using MCMC to explore parameters in the model with $m_{\phi_1}$ fixed
at 3 GeV, we find that the branching fraction to muons, $f_\mu$, is
concentrated near 0.9 for fits to Fermi, while having a broader
distribution in CCW.  This is shown in fig.\ \ref{fig:two-med2}(left).
The best-fit values of the annihilation cross section $\langle\sigma
v\rangle$ are shifted upwards by a factor of a few in these models
relative to the single-mediator model, as can be seen from fig.\
\ref{fig:two-med2}(right) in comparison to fig.\ 
\ref{fig:degen} (bottom right).  Here we also show constraints on
annhilations from Fermi dwarf observations (see discussion of fig.\
\ref{fig:degen}), and from our analysis of antiproton constraints,
for the representative case of 20\% branching fraction to $b$ quarks,
except for the latest Pass 8 constraint, where we use 10\% to 
$b$ quarks.  There is strong tension between this limit and the
preferred region for the GC excess of the Fermi data, while some
compatibility with the CCW data remains.
The antiproton limit on the cross section is relaxed by a factor 
of 5 compared to the limit for annihilations purely into $b$s
shown in fig.\ \ref{fig:AntiProton}.  The cross sections needed for
the GeV excess are still comfortably below this limit (and 
more so in the single-mediator model where the target value of the
cross section is lower).

\subsection{Sommerfeld enhanced annihilation}
\label{sommerfeld}

The fact that these cross sections are significantly higher than 
the nominal value $\langle\sigma v\rangle_0 \equiv 3\times
10^{-26}$cm$^3/s$, needed for approximately the right relic density, is potentially
a cause for
concern.  Such large cross sections would significantly suppress the 
density
of $\chi$, in contradiction to our assumptions.  However the presence
of the light mediator $\phi_1$ provides a possibility for resolving
this problem, due to the low velocities of DM in the galaxy relative
to the early universe, and the resulting nonperturbative
Sommerfeld enhancement of
the cross section by multiple exchanges of $\phi_1$.  

Let us suppose that $\phi_1$ couples to $\chi$ with strength $g_1$
and define $\alpha_1 = g_1^2/4\pi$. (Later we will see that
both scalar $g_{1}$ and pseudoscalar $g_{1,5}$ couplings are needed 
to get $s$-wave annihilation, but that $g_{1,5}\ll g_1$.)  The Sommerfeld enhancement $S$
is controlled by the two small parameters 
\cite{ArkaniHamed:2008qn}
\be
	\epsilon_{\phi} = {m_{\phi_1}\over \alpha_1 m_\chi},\quad
	\epsilon_v = {v\over\alpha_1}
\ee
A good approximation to $S$ is given by
the expression 
\cite{Cassel:2009wt,Slatyer:2009vg} 
\be S = \left({\pi\over \epsilon_v}\right){\sinh
X \over \cosh X - \cos\sqrt{{(2\pi/\bar\epsilon_\phi)} - X^2}}
\label{Seq} \ee where $\bar\epsilon_\phi= (\pi/12)\epsilon_\phi$ and $X
= \epsilon_v/\bar\epsilon_\phi$.  (The cosine becomes $\cosh$ if the
square root becomes imaginary.)  

\begin{figure*}[t]
\centerline{
\includegraphics[width=1\columnwidth]{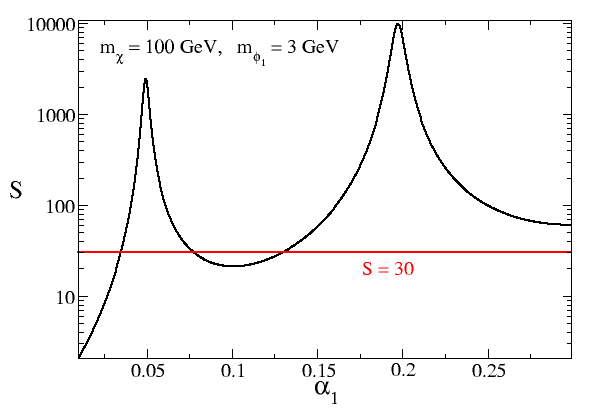}
\includegraphics[width=1\columnwidth]{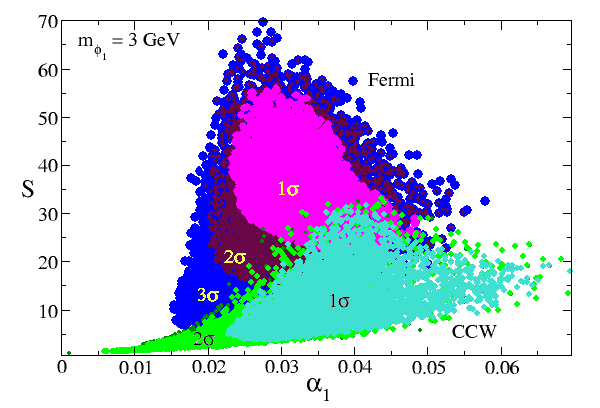}
}
\caption{Left: example of Sommerfeld enhancement factor as a function
of coupling $\alpha_1 = g_1^2/4\pi$ of light mediator $\phi_1$ to
dark matter, with $m_\chi=100$ GeV and $m_{\phi_1}=3$ GeV.  Typical
required value of $S=30$ is shown by the horizontal line, for
reference.  Right:
scatter plot of required values of $S$ versus minimum value of
$\alpha_1$ corresponding to sample of models in fig.\ \ref{fig:two-med2}.}
\label{fig:S}
\end{figure*}

To quantify the magnitude of Sommerfeld enhancement needed, and the
corresponding coupling strengths $\alpha_1$, for each model in our
MCMC chains we estimate the needed value of $S$ as the actual cross
section $\langle\sigma v\rangle$ divided by the nominal  relic density
value $\langle\sigma v\rangle_0$, assuming that $v=10^{-3}$.  
Typically there are several coupling strengths that can yield the
required value of $S$ due to resonances, illustrated in fig.\
\ref{fig:S}(left). Scanning over $\alpha_1$, we find the minimum value
of the coupling needed to get the required enhancement.  The
distributions of $S$ versus $\alpha_1$ for fitting the Fermi and CCW
data are shown in fig.\  \ref{fig:S}(right).  It is noteworthy that
we can get the right amount of enhancement by invoking reasonably small
values of the couplings, with $0.01\lesssim\alpha_1\lesssim 0.06$.

\begin{figure*}[t]
\centerline{
\includegraphics[width=1\columnwidth]{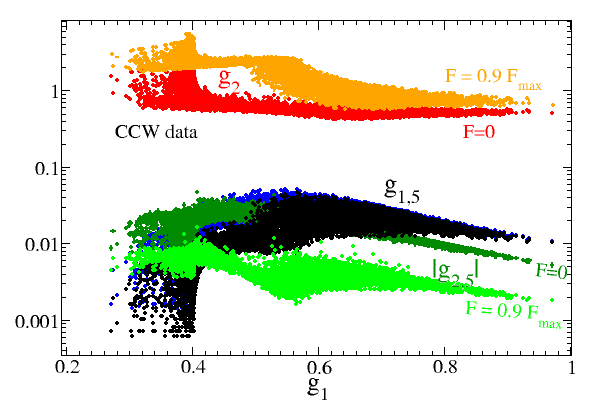}
\includegraphics[width=1\columnwidth]{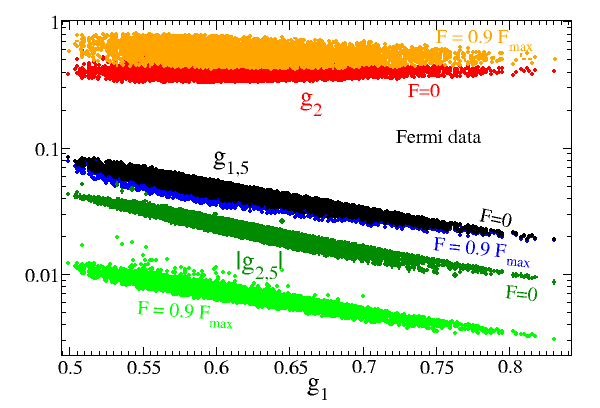}
}
\caption{$2\sigma$-allowed values of mediator couplings to DM for
fits to CCW (left) and Fermi (right) data sets.  Regions differ
depending upon the choice a one free combination of couplings which we
take to be $F$ (eq.\ (\ref{Feq}), which we illustrate for $F=0$
and $F=0.9\,F_{\rm max}$ (see eq.\ (\ref{Fmaxeq})).}
\label{fig:couplings}
\end{figure*}

\section{Particle physics models}
\label{models}

To further explore the implications of dark matter with two scalar
mediators, such as from  the thermal relic density requirement and
direct detection constraints, we need to more fully specify the 
interactions.  We will allow for parity-breaking in the dark sector,
with both scalar and pseudoscalar couplings
\be
\label{couplings}
	{\cal L}_{\rm int} = \sum_{i=1}^2 \bar\chi 
	\phi_i(g_{i} + i g_{i,5}\gamma_5)\chi
\ee
This choice is meaningful if the DM mass $m_\chi$ is 
restricted to be
real; otherwise $g_{i,5}$ can be rotated away by a chiral redefinition
$\chi\to e^{i\theta\gamma_5}\chi$. The ansatz (\ref{couplings})
is motivated by the fact that the cross section for 
$\chi\bar\chi\to\phi_i\phi_j$
is $p$-wave suppressed unless both types of couplings are present.
For simplicity we take both mediators to be real fields.

However, both scalars are required to get VEVs in order to mix with
the SM Higgs and hence couple to SM fermions, so in fact eq.\ 
(\ref{couplings}) would lead to a complex $m_\chi$ once the VEVs
are taken into account, and we would have to redefine the fields to
make it real.  Instead we will regard (\ref{couplings}) as describing
the low-energy effective theory where this has already been carried
out.  

The other important parameters of the model are the mixing angles
$\theta_i$ between the mediators and the SM Higgs.  These are
determined by cross-couplings $\frac12\lambda_i \phi_i^2 |H|^2$ and
VEVs,
\be
	2\theta_i \cong {\lambda_i \langle\phi_i\rangle v
	\over m_h^2 - m_{\phi_i}^2}
\label{theta_eq}
\ee
in the limit of $\theta_i\ll 1$, where $v = \sqrt{2}\langle H\rangle
=246$ GeV.
For our purposes, knowledge of the individual parameters $\lambda_i$
and $\langle\phi_i\rangle$ is not necessary, and we will work directly
with the $\theta_i$, that lead to couplings of the mediators to
SM fermions $\psi$ of the form
\be
	\theta_i y_j \phi_i \bar \psi_j \psi_j
\ee
where $y_j = m_j/v$ is the SM Yukawa coupling.

For simplicity we have neglected the mediator mixing mass
$m_{12}^2\phi_1\phi_2$ and the cross-coupling
$\lambda_{12}\phi_1^2\phi_2^2/4$.   They get generated at one loop  by
a virtual Higgs or $\chi$, at the level $\lambda_{12} \sim 
(\lambda_1\lambda_2- g_1^2 g_2^2)/(16\pi^2)$.  Cubic terms
$\phi_1\phi_2^2$ and $\phi_2\phi_1^2$ can be forbidden by the 
discrete symmetry $\phi_i\to -\phi_i$, $\chi\to
e^{i\pi\gamma_5/2}\chi$ as long as $\chi$ gets all of its mass from
the singlet VEVs (in our case primarily $\langle\phi_2\rangle$). This
symmetry is spontaneously broken by the VEVs of $\phi_i$ so these
terms get generated at one loop.  They allow for the decay
$\phi_2\to\phi_1\phi_1$ which we have ignored.  Taking
it into account  will require some small shift in the values of
$g_1/g_2$ needed to get the desired ratio of muons to $b$ quarks for
the GC excess signal, but since the data are not yet good enough to
determine these couplings precisely, we do not expect the
$\phi_2\to\phi_1\phi_1$ decay channel to change our results significantly.

To see how much fine-tuning is required by our neglect of
$\phi_1$-$\phi_2$ mass mixing, the most important term to consider is
the $\chi$-loop diagram connecting $\phi_1$ to $\phi_2$, and leading to $m^2_{12} \sim g_1g_2m_\chi^2/(16\pi)$.  If $m_1^2$ and $m_2^2$ denote the diagonal terms in the
mass matrix, the lightest eigenvalue is given by
\be
	m_{\phi_1}^2 \cong m_1^2 - {m_{12}^4\over m_2^2}
\label{mphi1}
\ee
Taking $m_2\sim m_\chi\sim 100$ GeV, the second term is
of order $-1\,$GeV$^2$, similar in size to $m_1^2$ (we have assumed
$m_1 = 3\,$ GeV for our benchmark models).  Therefore no extraordinary
fine tuning seems to be required in our model to keep $m_1$ small
(of course we ignored here the Planck-scale hierarchy problem and
took the heaviest threshold in the hidden sector for the estimate of
the loop contribution).  An accidental cancellation between the
terms in (\ref{mphi1}) could even help to explain the smallness of
of $m_1$.

\section{Relic Density}
The relic density of $\chi$ is determined by $\chi\bar\chi \to
\phi_i\phi_j$ summed over all possible final states.  At kinematic
threshold we find that 
\bea
	\sigma v_{\rm rel} &\cong& \sum_{i=1,2}
	\frac{g_i^2 g_{i,5}^2 m_\chi \sqrt{m_\chi^2-m_i^2}}
	{8 \pi  \left(
   m_\chi^2-m_i^2/2\right)^2}\nonumber\\
	&+&
{F^2\over 16\pi (m_\chi^2 - m_2^2/4)}
\label{sveq}
\eea
where for simplicity we approximated $m_1=0$ in the second 
line, and defined
\be
	F = (g_1 g_{2,5} + g_2 g_{1,5}) 
 + (g_1 g_{2,5}-g_2 g_{1,5}){m_2^2\over
	4 m_\chi^2}
\label{Feq}
\ee

The couplings are constrained by the fits to the GC excess indicating
that the number of muons to $b$ quarks produced in the annihilations
is given by
\be
	{N(\mu)\over N(b)} = {N(\phi_1)\over N(\phi_2)} \sim 
	{(g_1 g_{1,5})^2 + F^2/4\over (g_2 g_{2,5})^2 + F^2/4}
\label{Req}
\ee
For simplicity we here omitted the dependence upon $m_2^2/m_\chi^2$
implied by eq.\ (\ref{sveq}), but the more exact expression is
given in appendix \ref{appB} (eq.\ (\ref{Req})).

By demanding that $\langle\sigma v\rangle$ matches the canonical cross
section $\langle\sigma v\rangle_0 = 3\times 10^{-26}$ cm$^3$/s
for approximately achieving the observed relic density, and 
also that (\ref{Req}) gives the required ratio of $\mu/b$
for a given model realization, we get two constraints on the 
couplings.  The requirement of sufficient Sommerfeld enhancement
gives a third, fixing the magnitude of $g_1$.  We can take the value
of $F$ in (\ref{Feq}) as a free parameter that can be varied to 
explore the range of possible solutions for $g_i,g_{i,5}$
characterizing viable models.  Details of the algebraic solution
for the couplings are given in appendix \ref{appB}. 

We have carried out the above procedure for the models in our Monte
Carlo searches that give good fits to the GC GeV excess, to find the
ranges of allowed values for the couplings.  Scatter plots of 
$g_{1,5}$, $g_2$, $g_{2,5}$ versus $g_1$ are shown in fig.\ 
\ref{fig:couplings}
for fits to the Fermi and CCW data sets, within the $2\sigma$-allowed
regions.  Here we have assumed solutions for $g_i$ that give the
smallest values of the couplings (since a quadratic equation must be
solved leading to a second branch of solutions with larger values).
Interestingly, the parity-conserving couplings turn out to be the
largest, with $g_1\sim g_2\sim (0.5-0.8)$, while the parity-violating 
couplings are suppressed, with $g_{1,5}\sim (0.02-0.1)$ and $|g_{2,5}|$
being smaller.  The exact range of $g_{2,5}$ (which we take to be
negative, see appendix \ref{appB}) is the least constrained of all the
couplings, with $g_{2,5}=0$ always being a possibility (corresponding 
to $F = F_{\rm max}$), since we have
freedom to impose this as an extra condition while satisfying the
remaining physical constraints.  But in no case can $|g_{2,5}|$ be large
while maintaining that $\mu$ final states dominate over $b$ for the
GeV excess spectral shape.

\section{Direct detection}

\begin{figure}[b]
\centerline{
\includegraphics[width=1\columnwidth]{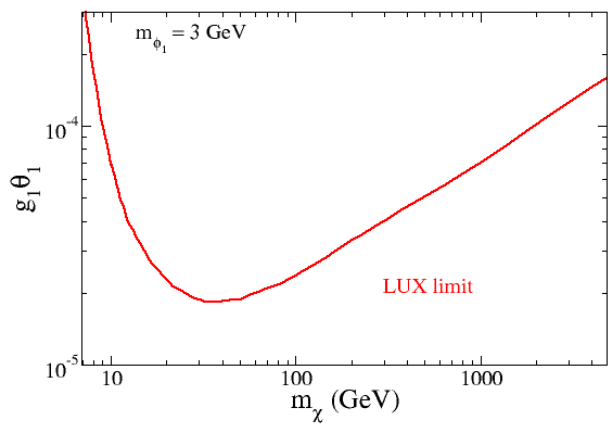}
}
\caption{Upper limit on DM coupling times mixing angle of the
light mediator $\phi_1$ from LUX \cite{Akerib:2013tjd} direct search.}
\label{fig:lux-limit}
\end{figure}

The cross section for scattering on nucleons (with mass $m_p$) 
is dominated by exchange of the light $\phi_1$ mediator,
\be
	\sigma_p \cong {(g_1\,\theta_1\, y\, m_p)^2\over
	\pi\, m_{\phi_1}^4}
\ee
in the limit that $m_\chi\gg m_p$.  Direct detection limits put an 
upper bound on $g_1\theta_1$ that depends upon $m_\chi$ as shown in fig.\ \ref{fig:lux-limit} for 
fixed $m_{\phi_1} = 3$ GeV.  Here $y$ is the Higgs-nucleon coupling,
which we take to be $y=1.3\times 10^{-3}$ (see ref.\ \cite{Cline:2013gha}  for
a recent review).  For $m_\chi\sim 100$ GeV as suggested by our fits
to the GC excess, this gives $g_1\theta_1 < 2\times 10^{-5}$.
On the other hand the Sommerfeld enhancement determined in section
\ref{sommerfeld} demands that $\bar g\sim 0.6$, so $\theta_1 \lesssim
3\times 10^{-5}$.  Eq.\ (\ref{theta_eq}) then implies $\lambda_1
\langle\phi_1\rangle < 4\times 10^{-3}$ GeV.  We can eliminate
the singlet VEV using its relation to 
the $\phi_1$ mass and self-coupling
$\bar\lambda_1$ by $m_{\phi_1} = \sqrt{2}\bar\lambda_1
\langle\phi_1\rangle$.  Taking $m_{\phi_1}=3$ GeV, we see that the 
direct detection constraint implies a hierarchy between the scalar
quartic couplings,
\be
	{\lambda_1\over \bar\lambda_1^{1/2}} < 2\times 10^{-3}
\ee
This is not technically unnatural since the coupling $\lambda_1$
only receives multiplicative renormalizations.

\section{Cosmological Constraints}

We consider here the impact of cosmic microwave background
constraints, and those coming from scattering of 
dark matter on itself in this section.  We will show that the models
treated in this work are compatible with current data, but are close
to the limits, with interesting potential to address puzzles in the
small scale structure of galaxies predicted by noninteracting cold dark matter.

\subsection{Cosmic microwave background}

Dark matter annihilations into charged particles during the epoch of
reionization can impact the temperature and polarization fluctuations
of the cosmic microwave background (CMB)
\cite{Chen:2003gz}-\cite{Madhavacheril:2013cna}, leading to contraints on
the annihilation cross section that are particularly strong for low DM
masses, scaling linearly with $m_\chi$. At this late era, the DM velocity is already sufficiently low
that the magnitude of Sommerfeld enhancement occuring in collisions at
the galactic center will be operative also for the CMB.  Therefore we
can directly compare the cross sections needed for the GC excess to
the CMB limits.  

The CMB limits are depend upon the efficiency $f_{\rm eff}$ for 
given final state particles to deposit energy electromagnetic energy
in the plasma.  The latest upper limit on the annihilation cross section 
using Planck temperature and polarization data can be expressed as
\be
	f_{\rm eff}\langle\sigma v\rangle < 4\times
10^{-26}\left(m_\chi\over \rm 100\ GeV\right){\rm cm^3/s}
\ee
which we infer from fig.\  41 of ref.\ \cite{Planck:2015xua}.  Ref.\ 
\cite{Cline:2013fm}
gives $f_{\rm eff} = 0.25\,(0.33)$ for $\mu\,(b$) final states at
$m_\chi = 100$ GeV, which we average to $f_{\rm eff} = 0.27$ for 
our benchmark two-mediator model with $80\%\,\mu+20\%\,b$ final states,
to give $\langle\sigma v\rangle< 1.5\times 10^{-25}(m_\chi/100{\rm\
GeV})\,{\rm cm^3/s}$.  This is plotted on fig.\ \ref{fig:two-med2}
(short dash-dotted curve).
It is a somewhat weaker constraint that the latest (Pass 8) limit from
Fermi dwarf galaxy observations, but still excludes the
preferred  regions for the GeV excess fits to Fermi data with
two-mediator models, while remaining marginally compatible with
the fits to the CCW data.  

\subsection{Dark matter elastic self-interactions}

Self-interactions of dark matter can be significant in our
two-mediator model, from $t$- and $u$-channel exchange of the lighter
mediator.  The viscosity cross section, relevant for effects
of DM self-scattering on structure formation, is\footnote{for
scattering of identical particles, this is more appropriate than
weighting by $(1-\cos\theta)$ since it treats scattering by
180$^\circ$ as equivalent to forward scattering}
\be
	\sigma_v =  \int d\Omega\,(1-\cos^2\theta)\,{d\sigma\over d\Omega} = 
	{2\, g_1^4\, m_\chi^2\over 3\pi\, m_{\phi_1}^4}
\ee

Taking the typical values $g_1\sim 0.6$, $m_{\phi_1} = 3$ GeV,
$m_\chi= 100$ GeV, indicated by our fits, this leads to a cross section
per DM mass of order $\sigma_v/m_\chi\sim 10^{-5}\,$b/GeV, far below
the bound $\sim 0.5\,$b/GeV from simulations of structure formation 
including DM self-interactions \cite{Zavala:2012us}.   However, there
is freedom in our model to take $m_{\phi_1}$ as small as $2 m_\mu = 
0.2\,$ GeV, if we adjust $\alpha_1$ to somewhat smaller values 
$\sim 0.01$ to compensate for the increased boost from Sommerfeld
enhancement in the galactic center.

The self-interaction cross section in such models can have
nonperturbative enhancements in analogy to the Sommerfeld effect,
which have been studied in detail in  ref.\
\cite{Tulin:2012wi,Tulin:2013teo}.  For $\alpha_1=0.01$,  the latter
reference finds that $\sigma_v/m_\chi$ is at the right level to
address problems of collisionless cold dark matter for predicting the
observed galactic small scale structure, if $m_\chi\cong 165\,
m_{\phi_1}$ in the  region $m_{\phi_1} = 0.2-2$ GeV, which  overlaps
with values needed for explaining the GC excess. These problems
include the difficulty for  collisionless cold dark matter to
correctly predict abundances and maximum masses of dwarf satellite
galaxies, as well as the cusp versus core issue for dwarf galaxy DM
density profiles; see ref.\  \cite{Weinberg:2013aya} for a review.

\section{Collider Constraints}

We also consider the collider limits on the particle physics model
described in section \ref{models}. In a model having additional scalars,
which mix with the SM Higgs, one would expect possible limits
resulting from the recent observation and measured signals of the
Higgs boson at the LHC. Although we find the limits posed by direct
detection and relic density results to be more constraining, we
include here the collider considerations for completeness. 

For the following anaysis, we do not initially assume a small mixing angle approximation. We supplement the SM Higgs potential with two real, scalar, singlet fields. Both fields acquire vevs, thereby inducing mixing with the Higgs. The relevant parameters are then the masses of the three scalars, $m_h, m_{\phi_1}, m_{\phi_2}$, and the two mixing angles, $\theta_{1h}, \theta_{2h}$. Mixing between the field and mass eigenstates is given according to 

\begin{equation}
\begin{pmatrix} \tilde{h} \\ \tilde{\phi}_1 \\  \tilde{\phi}_2 \end{pmatrix} = 
\begin{pmatrix} c_{\theta_{1h}}c_{\theta_{2h}} & s_{\theta_{1h}} & c_{\theta_{1h}}s_{\theta_{2h}} \\  
                        -s_{\theta_{1h}} c_{\theta_{2h}}& c_{\theta_{1h}} & -s_{\theta_{1h}}s_{\theta_{2h}} \\
                         -s_{\theta_{2h}} & 0 & c_{\theta_{2h}} \end{pmatrix} 
\begin{pmatrix} h \\ \phi_{1} \\ \phi_{2} \end{pmatrix}
\end{equation}
where $(c,s)$ denote $(\cos,\sin)$, respectively.

\begin{figure}[b!]
    \includegraphics[width=\linewidth]{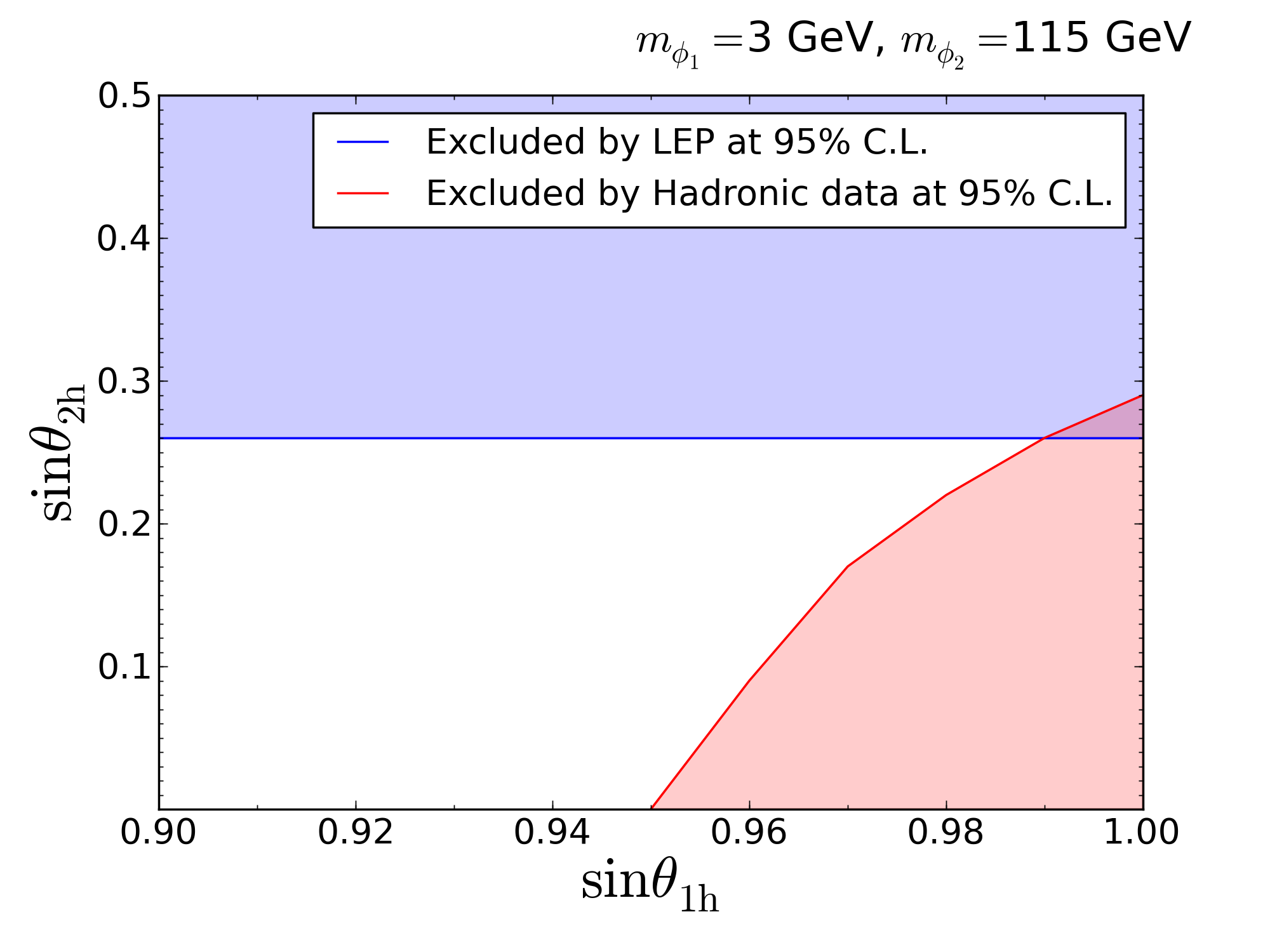}
    \includegraphics[width=\linewidth]{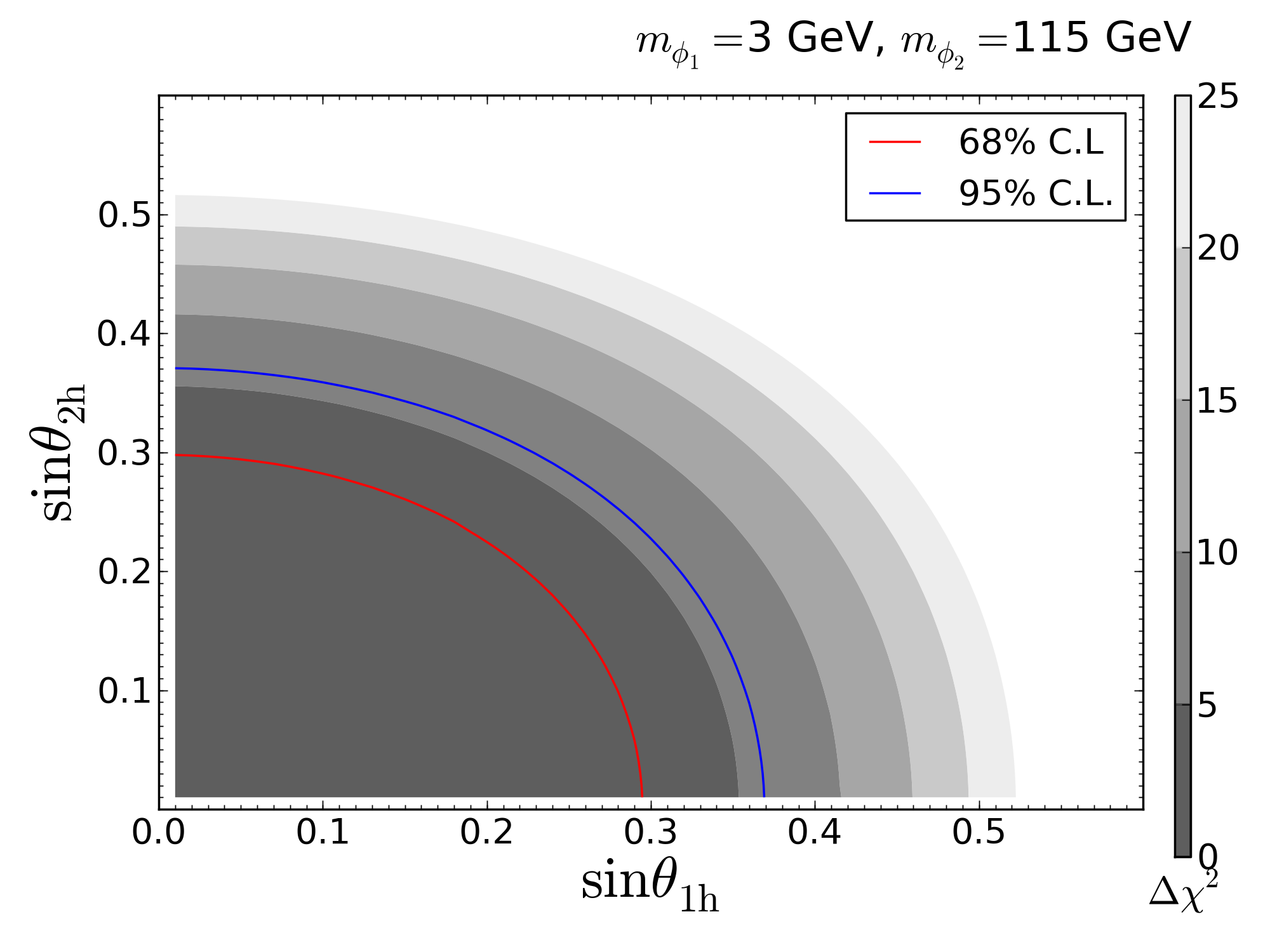}
  \caption{Excluded regions of $(\theta_{1h}, \theta_{2h})$ parameter space, under collider constraints. 
Top: blue (red) shaded region is excluded at 95\% c.l. by LEP (LHC/Tevatron) exclusion limits. Bottom: preferred regions, compatible with LHC Higgs signal strength measurements. Red and blue curves correspond to 68\% and 95\% c.l.}
\label{fig:fig1}
\end{figure}

The scalar couplings to standard model particles are then simply the SM Higgs couplings, scaled by $c_{\theta_{1h}}
c_{\theta_{2h}}$, $s_{\theta_{1h}}$, $c_{\theta_{1h}}s_{\theta_{2h}}$,    for $h$, $\phi_1$, and $\phi_2$
respectively.  We take $h$ to be the recently discovered Higgs boson, setting $m_h = 125.6$  GeV, and take $\phi_1$ to
be the lighter scalar, such that $m_{\phi_1} < m_{\phi_2}$. We determine the allowed values of the two mixing angles,
using the publicly available code HiggsBounds 4.2.0
\cite{Bechtle:2008jh,Bechtle:2011sb,Bechtle:2013gu,Bechtle:2013wla} and HiggsSignals 1.3.0
\cite{Stal:2013hwa,Bechtle:2013xfa}. We fix the values of the masses, such that $m_h$ agrees with the discovered
Higgs, as discussed above, and the values of $m_{\phi_i}$ are those preferred by fits to the GC excess, $m_{\phi_1} =
3$ GeV and $m_{\phi_2} = 115$ GeV. 

The result is shown in fig. \ref{fig:fig1}. The upper figure shows the regions excluded by LEP and hadronic (Tevatron
and LHC) exclusion Higgs searches, obtained by HiggsBounds, while the lower one  shows the regions preferred by
compatibility with the observed Higgs signal strengths. The set of experimental results and specific Higgs channels
that we use for the Higgs signal limit, can be found in refs
\cite{Aad:2014aba,ATLAS-CONF-2014-061,ATLAS-CONF-2014-060,Aad:2014eva,Aad:2014xzb,
Khachatryan:2014ira,Chatrchyan:2014vua,Chatrchyan:2013iaa,Chatrchyan:2013mxa,CMS:yva}. For a complete list of the
results used in the exclusion bounds analysis, see \cite{Bechtle:2013wla} and references therein.

\begin{figure}[tb]
\centering
\includegraphics[width=0.5\textwidth]{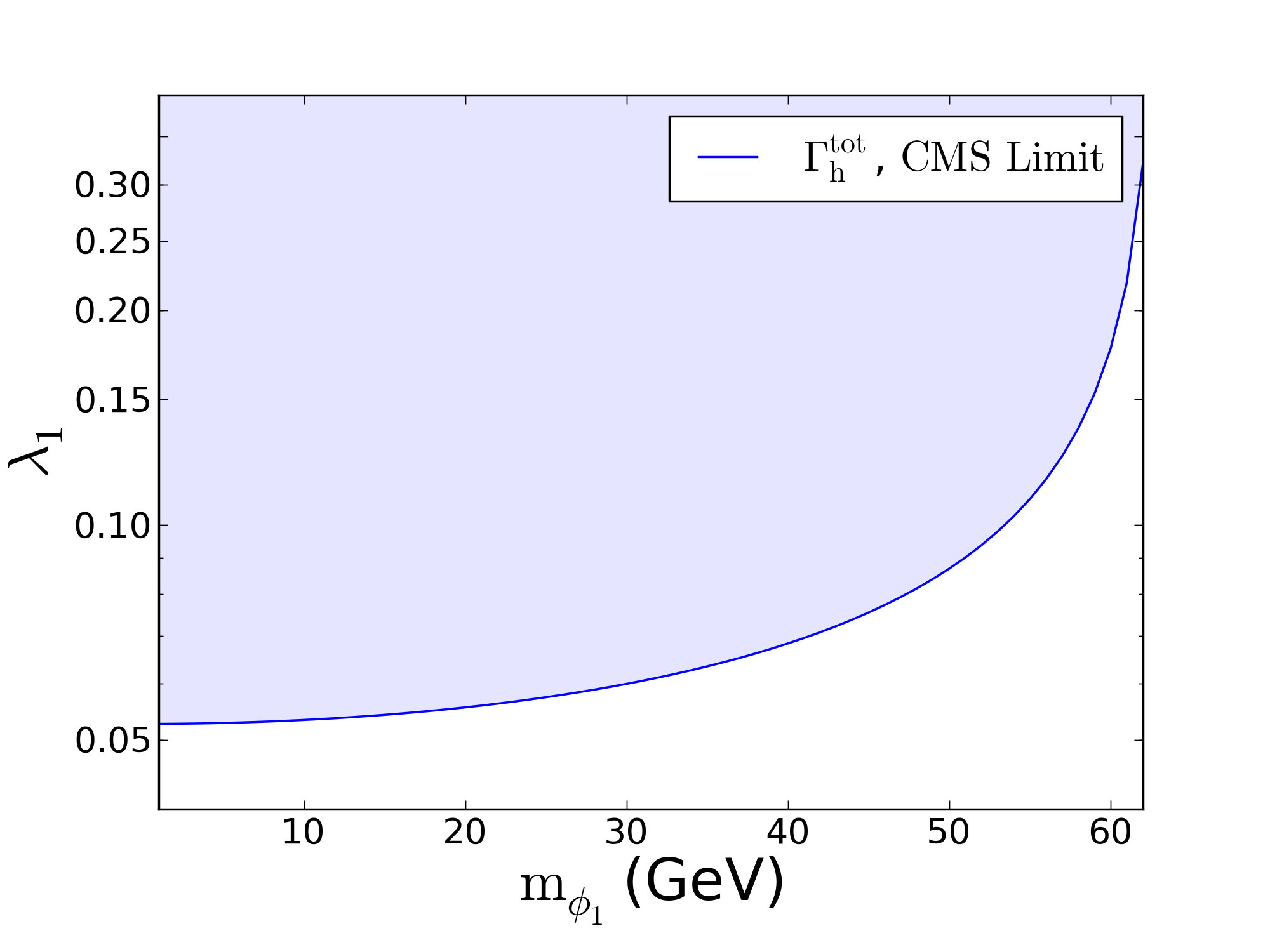}
\caption{Upper limit on $\phi_1$-Higgs cross coupling vs. $m_{\phi_1}$, from CMS constraint on the 
total Higgs width, at 95\% c.l. 
The shaded region is excluded.}
\label{fig:fig2}
\end{figure}

In the described analysis of Higgs signal limits, we only consider the effect of the mixing angles on scaling of the scalar couplings. In principle, new contributions to the width may further suppress the branching ratio, and signal rate, in some channels. For simplicity, we present the results taking only SM decays of the scalars, as the collider limits turn out to be much less constraining than other limits considered here.

For the mass hierarchy that is suggested by the GC considerations, one must also consider the additional contributions to the Higgs width, arising from decays to the light scalar, $h \rightarrow \phi_1 \phi_1$.  Returning to the small mixing approximation, the new contribution to the width is given by
\begin{equation}
\Gamma_{h \rightarrow \phi_1 \phi_1} = \frac{\lambda_1^2 \nu^2}{32 \pi m_h} \left( 1 - \frac{4m_{\phi_1}^2}{m_h^2}. \right)^{1/2}
\end{equation}

A recent CMS analysis \cite{CMS:2014ala} presents an upper limit on
the Higgs total width, obtained from the ratio of measurements of
off-shell and on-shell Higgs production and decay in the $H
\rightarrow Z Z$ channel. CMS finds $\Gamma_h < 4.2\, \Gamma_h^{SM}$, with $\Gamma_h^{SM} = 4.15$ MeV. We find the upper limit on the cross coupling shown in fig. \ref{fig:fig2}.

\section{Conclusions}

In this study we compared two new data sets,\footnote{in particular,
the Fermi collaboration spectrum determined within energy bands,
as opposed to their ansatz using a power law with exponential cutoff,
has not previously been analyzed with respect to dark matter models}
 characterizing the excess
gamma rays from the galactic center, to predictions from  models where
dark matter annihilates into light mediators that subsequently decay
into standard model particles.  In contrast to models with direct
annihilation into heavy quarks, these are more easily
compatible with direct detection and cosmic ray antiproton
constraints, as we have demonstrated. In a first approach, general
admixtures of final state particles yielded good fits, revealing a
preference for the data to be described by decays mostly to muons
(electrons being excluded by AMS-02 data)  and $b$ quarks, with dark
matter masses in the range $40-140$ GeV. Either Majorana or Dirac dark
matter are viable possibilities.

We then argued that this might be more naturally  accomplished with
two scalar mediators $\phi_i$ mixing with the Higgs, such that
$\phi_1$ decays primarily to $\mu^+\mu^-$ and $\phi_2\to b\bar b$, as
a consequence of the mediator masses.  Encouragingly, fits to the data
in this two-mediator model were consistent with masses in the desired
ranges $2m_\mu < m_{\phi_1} < 2 m_\tau$, $2m_b < m_{\phi_2} < 
m_\chi$.  Moreover quite reasonable perturbative values of the
mediator couplings to $\chi$ give consistent results.  To avoid
$p$-wave suppression of  the annihilation cross section we invoked
parity violating couplings, which turn out to be somewhat smaller than
the parity-conserving ones.

An interesting feature of these models is that the best-fit cross
sections for annihilation in the galactic center are several times
larger than the value needed for achieving the right thermal relic
density.  We showed that this can be consistent due to Sommerfeld
enhanced annihilation in the galaxy, due to multiple exchanges of the
lighter of the two mediators, taking $m_{\phi_1} = 3$ GeV as a
benchmark model.  It only requires that the scalar coupling of
$\phi_1$ to $\chi$ be of order $g_1\sim 0.5$.  If the light mediator
mass is somewhat lower, the cross section for
elastic DM self-scattering can have the right magnitude
for addressing the missing satellite and cusp-core problems from
simulations of structure formation.  Improved CMB constraints
anticipated from Planck data may be in tension with the best fit 
regions of parameter space.

LHC constraints on the couplings and mixing angles of the mediators
are relatively weak compared to those from direct detection.  The
latter provides good prospects for independent confirmation of 
our model, requiring that the cross-coupling between $\phi_1$ and the
Higgs boson be $\lesssim 10^{-3}$.  No fine tuning in the technical
sense is needed to satisfy this constraint, but neither is there any
symmetry reason for the coupling to be small.  

The allowed parameter space in $\langle\sigma v\rangle$ versus
$m_\chi$ shows some tension with the latest Fermi-LAT constraints on
dark matter annihilation in dwarf spheroidals, especially for the fit
to the Fermi GeV excess data.  Planck constraints on distortions to
the CMB are in tension at a similar level.
The fit to the CCW excess is also
challenged by these results, though to a lesser extent.  More optimisitically, as we
were completing this work, preliminary evidence for a positive signal
from the dwarf galaxy Reticulum II appeared \cite{Hooper:2015ula}, at
a level consistent with expectations from the GC excess.

{\bf Acknowledgments.}  We thank S.\ Murgia for helpful correspondence
about the Fermi GeV excess spectrum, and M.\ Reece for stimulating
discussions.   J.C.\ is supported by the Natural
Sciences and Engineering Research Council (NSERC) of Canada.
The work of Z.L.\ is supported by the Tsinghua University Funds (under Grant No. 543481001 and
Grant No. 523081007).

\begin{table}[!htb!]
 \centering
\begin{tabular}[t]{|c|c|c|c|}
\hline
   {$E_\gamma$  [GeV]  } & 
   $\d \Phi / \d E_\gamma \d \Omega$ & 
   $\sigma_{\rm stat}$ & 
   $\sigma_{\rm syst}$ \\
\hline
\hline
1.122 & 1.587e-06 & 1.036e-07 & 8.225e-07 \\
\hline
1.413 & 1.624e-06 & 7.138e-08 & 5.810e-07 \\
\hline
1.778 & 1.483e-06 & 5.330e-08 & 3.240e-07 \\
\hline
2.239 & 1.122e-06 & 4.272e-08 & 1.226e-07 \\
\hline
2.818 & 7.298e-07 & 3.655e-08 & 5.857e-08 \\
\hline
3.548 & 4.265e-07 & 3.106e-08 & 4.964e-08 \\
\hline
4.467 & 2.475e-07 & 2.074e-08 & 3.511e-08 \\
\hline
5.623 & 1.405e-07 & 1.270e-08 & 2.735e-08 \\
\hline
7.079 & 7.662e-08 & 8.267e-09 & 1.874e-08 \\
\hline
8.913 & 4.039e-08 & 5.435e-09 & 1.226e-08 \\
\hline
11.220 & 2.272e-08 & 3.688e-09 & 7.959e-09 \\
\hline
14.125 & 1.345e-08 & 2.433e-09 & 4.936e-09 \\
\hline
17.783 & 7.828e-09 & 1.566e-09 & 3.016e-09 \\
\hline
22.387 & 4.341e-09 & 1.023e-09 & 1.820e-09 \\
\hline
28.184 & 2.503e-09 & 6.953e-10 & 1.115e-09 \\
\hline
35.481 & 1.600e-09 & 4.805e-10 & 6.589e-10 \\
\hline
44.668 & 1.029e-09 & 3.146e-10 & 4.090e-10 \\
\hline
56.234 & 5.832e-10 & 2.113e-10 & 2.782e-10 \\
\hline
70.795 & 2.753e-10 & 1.355e-10 & 1.556e-10 \\
\hline
89.125 & 9.287e-11 & 7.851e-11 & 6.110e-11 \\
\hline
\end{tabular}
\caption{Energy flux derived from Fermi collaboration's presentation  \cite{Murgia} 
for the galactic center gamma ray excess, and statistical along with
systematic errors as described in section \ref{fermi-spect}. 
The flux intensity is obtained by averaging the observed
total flux over 
the $15^\circ\times 15^\circ$ square around the GC.
 Flux units are $\mathrm{GeV^{-1} cm^{-2}
s^{-1} sr^{-1}}$.}
\label{table:fermi}
\end{table}

\appendix

\section{Fermi Spectrum}
\label{App0}
We list the Fermi spectrum in table \ref{table:fermi}.

\section{Spectrum from decay to muons}
\label{appA}

For completeness, we present the photon spectrum from 
$\phi \rightarrow \mu^+ \mu^-$ in the rest frame of $\phi$ 
\cite{Mardon:2009rc,Liu:2014cma}.
It includes photons from final state radiation and from 
radiative decays.
Final state radiation gives the contribution
\begin{equation}
 \frac{ \d N_{FSR}} {\d x} = \frac{\alpha_{em}} {\pi} \frac{ 1+ ( 1-x)^2} {x} 
     \left[ -1 +  \ln \left( \frac{m_\phi^2 ( 1-x)   } { m_\mu^2} \right) \right] \ ,
\label{mu1}
\end{equation}
where $x = 2 E_\gamma / m_\phi$.
The contribution from radiative decay is 
\begin{eqnarray}
   \frac{ \d N_{rad} } {\d x} &=& \frac{\alpha_{em}} { 3 \pi\,x} \bigg\{
      - \frac{17}{2} - \frac{3}{2} x + \frac{191}{12} x^2
      - \frac{ 23}{3} x^3
      + \frac{7}{4} x^4
   \nonumber\\
      &&
      + \left( 3 + \frac{2}{3} x - 6 x^2 + 3 x^3 - \frac{2}{3} x^4 + 5 x \ln x \right)
      \ln \frac{1}{r}  
   \nonumber\\
      &&
       + \left( 3 + \frac{2}{3} x - 6 x ^2 + 3 x^3 -
      \frac{2}{3} x^4 \right)  \ln ( 1-x)
   \nonumber\\
     &&
       - \frac{ 28}{3} x \ln x
         + 5 x \ln ( 1-x) \ln x + 5x \mathrm{Li}_2 (1-x)
      \bigg\}\nonumber\\
\label{mu2}
\end{eqnarray}
where $r = \frac{m_e^2}{m_\mu^2}  \ll 1 $, and the range of $x$ is $(0,1)$ which does not depends on $r$
since $r$ is negligible.
The sum of  (\ref{mu1}) and ({\ref{mu2}) gives us the total photon spectrum from annihilation to muons.

\section{Determination of couplings}
\label{appB}

Here we show how the couplings of the mediators to dark matter 
are analytically determined from the various observational
constraints.
We parametrize the annihilation cross section as
\be
	\langle\sigma v\rangle = ax + by + c F^2
\ee
where $x = (g_1 g_{1,5})^2$, $y= (g_2 g_{2,5})^2$, and $a,b,c$
are the functions of $m_\chi, m_{\phi_2}$ in (\ref{sveq}).  The
ratio $R$ of muons to $b$ quarks resulting from the annihilations is
\be
	R = {ax + c F^2/2\over by + c F^2/2}
\label{Req}
\ee
Setting the cross section equal to the relic density value 
$\langle \sigma v\rangle_0$ and using (\ref{Req}), we can solve 
for $x$ and $y$, for a given value of $F$:
\bea
	x &=& {1\over a}\left( {\langle \sigma v\rangle_0\over 1 +
	R^{-1}} -\frac{c}{2}F^2\right)\nonumber\\
	y &=& {1\over b}\left( {\langle \sigma v\rangle_0\over 1 +
	R} -\frac{c}{2}F^2\right)\nonumber\\
\eea
Since $x,y\ge 0$, this constrains
\be
	|F| \le F_{\rm max} = \left({2\, \langle \sigma v\rangle_0\over
		c}\right)^{\!\!1/2}\!\!{\rm min}\left[{1\over \sqrt{1+R}},\, 
	{\sqrt{R\over 1+R}} 	\right]
\label{Fmaxeq}
\ee

The minimum value of $\alpha_1$ needed for sufficient Sommerfeld
enhancement determines $g_1 =\sqrt{4\pi\alpha_1}$; then 
$g_{1,5} = \sqrt{x}/g_1$ and $g_{2,5} = \sqrt{y}/g_2$, where we should
allow for both possible signs of $\sqrt{x}$ and $\sqrt{y}$.
Eliminating $f_{1,5}$ and $f_{2,5}$ in equation (\ref{Feq}) for $F$ 
results in a quadratic equation for $s= g_1/g_2$:
\be
	s = {F \pm \sqrt{F^2 - 4\sqrt{x}\sqrt{y}(1-q^2)}\over
	2\sqrt{y}(1+q)}
\label{seq}
\ee
where $q = m_{\phi_2}^2/4m_\chi^2$.  The smallest value of $|s|$ is 
found when the signs of the two terms in the numerator of (\ref{seq})
are opposite.  We can take $F>0$ and the lower sign while keeping
$s>0$ if $\sqrt{y}<0$, corresponding to the choice of all couplings
except for $g_{2,5}$ being positive.  (Alternatively, we could take
all couplings except for $g_{1,5}$ positive and $F<0$, but there is no
physical difference.)   One can explore the range of possible couplings
by allowing $F$ to vary between 0 and $F_{\rm max}$.

\end{document}